\documentclass[intlimits,twoside,a4paper]{article}
\usepackage{wrapfig}
\usepackage{graphicx}
\usepackage{amssymb,amsmath}
\usepackage[T2A]{fontenc}
\usepackage[cp1251]{inputenc}

\usepackage[eqsecnum]{cmpj2}
\issue{2015}{18}{4}{43703}
\doinumber{10.5488/CMP.18.43703}

\newcommand{\be}{\begin{equation}}
\newcommand{\ee}{\end{equation}}
\newcommand{\bea}{\begin{eqnarray}}
\newcommand{\eea}{\end{eqnarray}}
\newcommand{\non}{\nonumber\\}

\title[Statistical theory of thermodynamic and dynamic properties of the RbHSO$_{4}$ ferroelectrics]%
{Statistical theory of thermodynamic and dynamic properties of the RbHSO$_{4}$ ferroelectrics}

\author[I.R. Zachek \textsl{et al.}]{I.R. Zachek\refaddr{label1}, R.R. Levitskii\refaddr{label2}, Ya.Shchur\refaddr{label2}, O.B.Bilenka\refaddr{label1}}
\addresses{
\addr{label1} Lviv Polytechnic National
University, 12 Bandera St., 79013 Lviv, Ukraine
\addr{label2} Institute for Condensed Matter Physics of the
National Academy of Sciences of Ukraine, \\ 1~Svientsitskii St.,
79011 Lviv, Ukraine }

\authorcopyright{I.R. Zachek, R.R. Levitskii, Ya.Shchur, O.B.Bilenka, 2015}
\date{Received June 10, 2015, in final form August 5, 2015}

\begin{document}

\maketitle

\begin{abstract}
Within the modified four-sublattice model of RbHSO$_{4}$ with
taking into account the piezoelectric coupling to the strains
$\varepsilon_i$, $\varepsilon_4$, $\varepsilon_5$, and
$\varepsilon_6$,  the polarization components, static and dynamic
dielectric permittivity of clamped and free crystal are calculated
in the mean field approximation. At the proper choice of the
values of the theory parameters, a satisfactory quantitative
description of the available experimental data is
obtained.
\pacs 77.22.Ch, 77.22.Gm, 77.65.-j, 77.80.Bh, 77.84.-s

\end{abstract}

\section{Introduction}

Chemical compounds such as sodium-potassium tartrate
NaKC$_{4}$H$_{4}$O$_{6}$$\cdot$4H$_{2}$O (Rs),
sodium-ammonium tartrate NaNH$_{4}$C$_{4}$H$_{4}$O$_{6}$$\cdot$4H$_{2}$O (ARs), rubidium
hydrosulphate RbHSO$_{4}$ (RHS), and ammonium hydrosulphate
NH$_{4}$HSO$_{4}$ (AHS)  belong to the order-disorder type
ferroelectrics. According to neutron and X-ray structure studies
of RHS \cite{nel,ash,ito,ito1}, Rb$^{87}$ \cite{kas1} and D$^{2}$
\cite{kas2,kas2a} NMR measurements, infrared  \cite{ mja} and Raman
scattering experiments \cite{baj}, the phase transition in RHS is
of  the second order. Protons are already ordered in the
paraelectric phase. Only one second-order phase transition point
($T_\textrm{c}=265$~K) is present. In the high-temperature phase, the
structure of RHS is monoclinic and is described by the space group P2$_{1}$/c--C$^{5}_\textrm{2h}$. Below the transition point, the monoclinic
symmetry  remains, but the space group changes to Pc--C$^{2}_\textrm{s}$.
The unit cell consists of eight molecules $Z = 8$ in both phases.

\begin{figure}[!h]
\begin{center}
\includegraphics[scale=0.4]{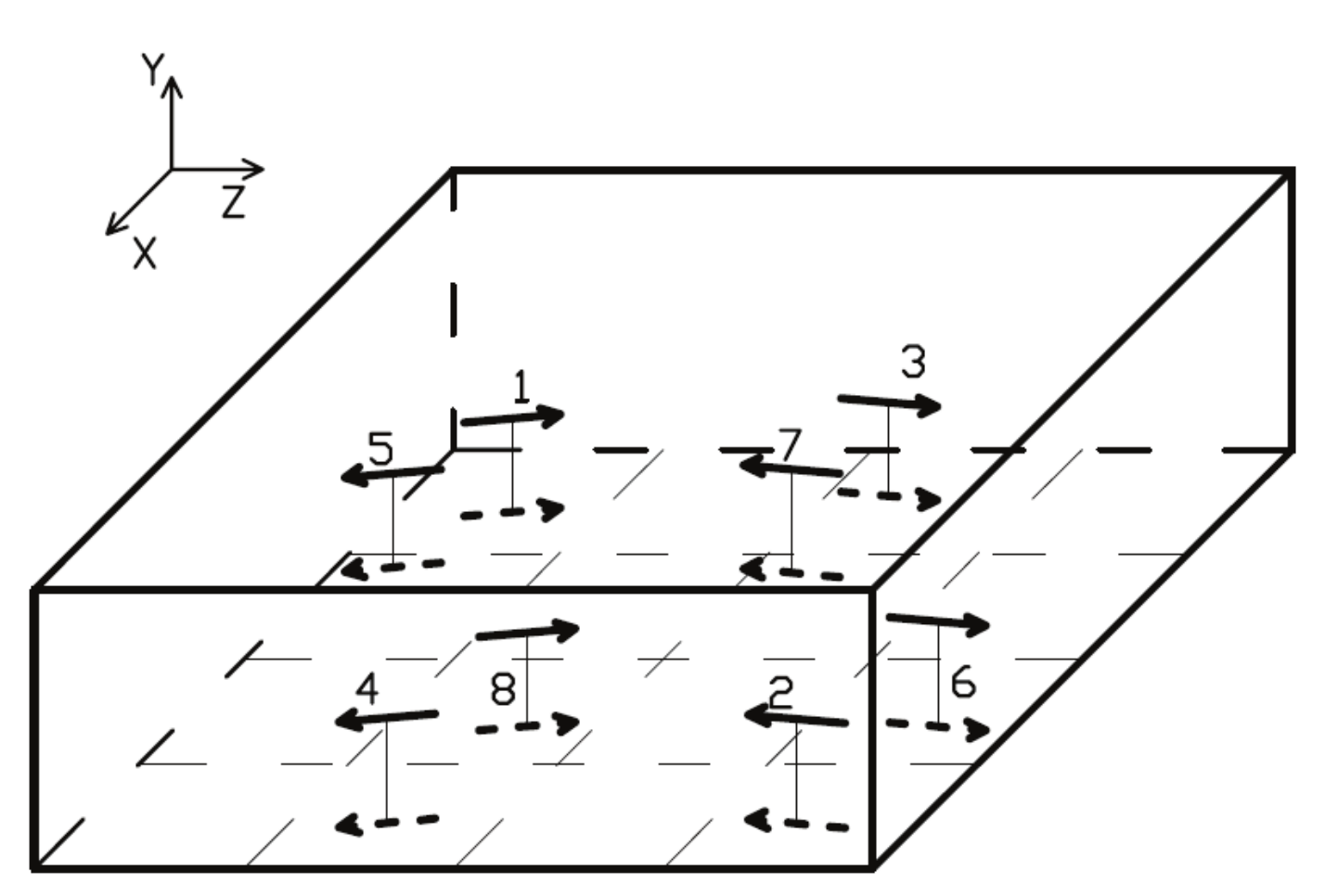}
\end{center}
\caption{Orientations of the ${\bf d}_{qf}$ vectors within the
primitive cell of  $R_s$ in the high-symmetry phase (the
paraelectric phase).} \label{paralelepiped}
\end{figure}

The phase transition in RHS is associated with the motion of sulphate
complexes (SO$_{4}$)$_{11}$, (SO$_{4}$)$_{12}$, (SO$_{4}$)$_{13}$,
(SO$_{4}$)$_{14}$ between two  equilibrium positions.
(SO$_{4}$)$_{25}$, (SO$_{4}$)$_{26}$, (SO$_{4}$)$_{27}$,
(SO$_{4}$)$_{28}$-complexes are completely ordered in the entire
temperature range and do not play any direct role in the
ferroelectric phase transition. These complexes form a  network of
elementary dipoles directed along the $z$-axis (figure~\ref{paralelepiped}).

The dipole moments created in the paraelectric phase by the (14)
and (12) (SO$_{4}$) complexes have the same direction which is
opposite to the direction of the  dipole moments created by the
(11) and (13) complexes. Analogously, the dipole moments of the
(25) and (26) type complexes are opposite to those of the (27) and
(28) complexes.

Two equilibrium positions (potential wells) of the ($1f$)-complexes
($f=1,2,3,4$) are not equivalent. Above the transition temperature
$T_\textrm{c}$, the $(1f)$-complexes are located in the energetically favorable
equilibrium positions (deeper potential wells). Mathematically,
non-equivalence of equilibrium positions is described by an
additional longitudinal field $\Delta$ acting on the dipoles of
the sulphate-complexes. This field is oppositely directed for the
(11), (14) and (12) complexes.

The phenomenological  \cite{nak} and statistic
\cite{bla,lev1,pap,gri,lev} models of the phase transition in
RHS-crystals,
 analogously to the  Mitsui model for Rs,  well describe the dielectric properties \cite{ ich, kaj,pep, nak1, fle}
  and the Debye-type dynamic dielectric permittivity \cite{ lev,lev1,pap,gri} in the mean
 field approximation. However, it is impossible to calculate
the experimentally measurable dielectric permittivity of a mechanically free crystal,
 the piezoelectric coefficients, elastic constants and transverse dielectric
permittivities  using these models
\cite{lev,lev1,pap,gri}.
That is why the piezoelectric coupling to the strains should be taken into account \cite{sta}.

In this work we propose a modification of the four-sublattice
model of the RbHSO$_{4}$ crystals, which takes into account the
piezoelectric coupling to the strains
 $\varepsilon_i$, $\varepsilon_j$ in the ferroelectric phase. The
 dielectric, piezoelectric, elastic, thermal and dynamic characteristics of RHS are calculated within the mean
 field approximation. The corresponding experimental data for this crystal are described.

\section{Four-sublattice model: Hamiltonian}

The system Hamiltonian is a modification of the Hamiltonian
proposed in \cite{stasjuk} that takes into account the
piezoelectric coupling, and a generalization of the Hamiltonian in
\cite{pap} to the 'three-dimensional' model. In the  quasi-spin
representation, it reads as follows:
\bea  \hat H &=&N U_\textrm{seed} - \frac12 \sum\limits_{qq'}\sum\limits_{f=1}^4 J_{ff}(qq')
\frac{\sigma_{qf}}{2}\frac{\sigma_{q'f}}{2} - \frac12
\sum\limits_{qq'}\sum\limits_{f\neq f'} K_{ff'}(qq')
\frac{\sigma_{qf}}{2}\frac{\sigma_{q'f'}}{2}  \nonumber\\
&& - \Delta \sum\limits_{q} \left( -\frac{\sigma_{q1}}{2} -
\frac{\sigma_{q2}}{2} + \frac{\sigma_{q3}}{2} +
\frac{\sigma_{q4}}{2}\right) -
 \mu_1E_1  \sum\limits_{q} \left(
 \frac{\sigma_{q1}}{2} - \frac{\sigma_{q2}}{2} +
\frac{\sigma_{q3}}{2} -
\frac{\sigma_{q4}}{2}\right)  \nonumber\\
&& - \mu_2E_2  \sum\limits_{q} \left(
- \frac{\sigma_{q1}}{2} + \frac{\sigma_{q2}}{2} +
\frac{\sigma_{q3}}{2} -
\frac{\sigma_{q4}}{2}\right)
 - \mu_3E_3  \sum\limits_{q} \left(
\frac{\sigma_{q1}}{2} + \frac{\sigma_{q2}}{2} +
\frac{\sigma_{q3}}{2} + \frac{\sigma_{q4}}{2}\right),  \label{eq:2.1}
\eea where $N$ is the  number of unit cells.
In (\ref{eq:2.1}) $J_{ff'}(qq')$ and $K_{ff'}(qq')$ are the
interaction potentials between the pseudospins from the same and
from different chains, respectively; the parameter $\Delta$
describes the asymmetry of the potential, in which the pseudospins
move; $\mu_i$ is the effective dipole moment per one pseudospin;
$\sigma_{qf}$ is the $z$-component of the pseudospin operator
situated  at $f$-th bond ($f=1,2,3,4$) in $q$-th unit cell.

$U_{1 \textrm{seed}}$ is the seed energy that includes the elastic,
piezoelectric and dielectric contributions expressed in terms of
 the electric fields $E_i$ $(i=1, 2, 3)$ and
strains $\varepsilon_i$ and $\varepsilon_j$ $(j=i+3)$.
$c_{jj}^{E0}(T)$, $e_{ij}^0$, $\chi_{ii}^{\varepsilon 0}$ are the
so-called seed elastic and piezoelectric constants and dielectric
permittivities:
\bea
U_\textrm{seed}&=& v
\bigg[\frac{1}{2}\sum\limits_{i,i'=1}^3c_{ii'}^{E0}(T)\varepsilon_i
\varepsilon_{i'}+
\frac{1}{2}\sum\limits_{j=4}^6c_{jj}^{E0}(T)\varepsilon_j^{2}-
 \sum\limits_{i=1}^3 e_{3i}^0 \varepsilon_i E_3 - e_{35}^0 \varepsilon_5 E_3
  \nonumber\\
&& \hspace{4mm}- \frac{1}{2}  \chi_{11}^{\varepsilon 0}E_1^2 - \frac{1}{2}
\chi_{22}^{\varepsilon 0}E_2^2 -
\frac{1}{2}  \chi_{33}^{\varepsilon 0}E_3^2\bigg],
 \eea
 $v$ is the unit cell volume.

Having analyzed the results of \cite{zaj}, we assume
the seed elastic constants $c_{jj}^{E0}(T)$ to be
linearly decreasing with temperature \cite{lis}
\bea
\label{eq:2.3}
  c_{ii'}^{E0}(T)= c_{ii'}^{E0}-k_{ii'}(T-T_\textrm{c}), \qquad c_{jj}^{E0}(T)=c_{jj}^{E0}-k_{jj}(T-T_\textrm{c}),
 \eea
where the coefficients $k_{jj}$ phenomenologically take into
account the high-temperature lattice anharmonism.

We make an identity transformation
%%% 2.2
\be \sigma_{qf} = \eta_f + (\sigma_{qf} - \eta_f), \qquad (f =
1,2,3,4), \ee
and neglect the quadric fluctuations. The Fourier
transforms of the interaction constants at ${\bf q}
 =0$, $J = J_{ff} = \sum\limits_{q'} J_{ff}(qq')$, $K_{ff'} = \sum\limits_{q'}
 K_{ff'}(qq')$ and $\Delta$  are expanded in series over the strains $\varepsilon_i$,
$\varepsilon_j$ up to the linear terms:
\begin{align}
 &J = J^0 + \frac{\partial J}{\partial \varepsilon_i}\varepsilon_i
 = J^0 + \sum\limits_{i=1}^3\psi_{1i}\varepsilon_i + \sum\limits_{j=4}^6\psi_{1j}\varepsilon_j, & K_{12} &=
 K_{12}^0 + \sum\limits_{i=1}^3  \psi_{2i}\varepsilon_i + \sum\limits_{j=4}^6\psi_{2j}\varepsilon_j,\nonumber \\
 &K_{13} = K_{13}^0 + \sum\limits_{i=1}^3
 \psi_{3i}\varepsilon_i + \sum\limits_{j=4}^6\psi_{3j}\varepsilon_j, &
 K_{14} &=
 K_{14}^0 + \sum\limits_{i=1}^3  \psi_{4i}\varepsilon_i + \sum\limits_{j=4}^6\psi_{4j}\varepsilon_j,\nonumber\\
&\Delta = \Delta^0 + \sum\limits_{i=1}^3\psi_{5i}\varepsilon_i + \sum\limits_{j=4}^6\psi_{5j}\varepsilon_j. &&
\end{align}

As a result, Hamiltonian (\ref{eq:2.1}) within the mean field approximation
takes the form:
\bea  H^{(0)} &=& v    U_\textrm{seed} + \frac18 J^{0}\left(\eta_1^2 + \eta_2^2 + \eta_3^2 + \eta_4^2\right) \nonumber\\
&&+\frac14 K_{12}^{0}\left(\eta_1\eta_2  +  \eta_3\eta_4\right)
  +  \frac14 K_{13}^{0}\left(\eta_1\eta_3  +  \eta_2\eta_4\right)  +
\frac14 K_{14}^{0}\left(\eta_1\eta_4  +  \eta_2\eta_3\right)\nonumber \\
&& + \frac18 \bigg(\sum\limits_{i=1}^3\psi_{1i}\varepsilon_i+ \sum\limits_{j=4}^6\psi_{1j}\varepsilon_j\bigg)\left(\eta_1^2 + \eta_2^2 + \eta_3^2 + \eta_4^2\right) +
\frac14  \bigg(\sum\limits_{i=1}^3  \psi_{2i}\varepsilon_i  +  \sum\limits_{j=4}^6\psi_{2j}\varepsilon_j\bigg)\left(\eta_1\eta_2  +  \eta_3\eta_4\right) \nonumber \\
&&  +  \frac14 \bigg(\sum\limits_{i=1}^3
 \psi_{3i}\varepsilon_i + \sum\limits_{j=4}^6\psi_{3j}\varepsilon_j\bigg)\left(\eta_1\eta_3  + \eta_2\eta_4\right)  +
\frac14 \bigg( \sum\limits_{i=1}^3  \psi_{4i}\varepsilon_i + \sum\limits_{j=4}^6\psi_{4j}\varepsilon_j\bigg)\left(\eta_1\eta_4  +  \eta_2\eta_3\right), \label{eq:2.7}\\
\hat H_s &=& - \sum\limits_q \left( {\cal H}_1
\frac{\sigma_{q1}}{2} + {\cal H}_2 \frac{\sigma_{q2}}{2} + {\cal
H}_3 \frac{\sigma_{q3}}{2} + {\cal H}_4 \frac{\sigma_{q4}}{2}
\right).
 \label{RReq2.5} \eea
Thus, we find the mean pseudospin values
%%% 2.7
\be \eta_f = {\tanh} \frac{\beta}{2}{\cal H}_f.
\ee
Let us use new variables:
%%% 2.8
\begin{align}
\xi_1 &= \frac14 ( - \eta_1 - \eta_2 + \eta_3 + \eta_4)
= \frac14 \left( {- \tanh} \frac{\beta}{2}{\cal H}_1 - {\tanh} \frac{\beta}{2}{\cal H}_2 + {\tanh} \frac{\beta}{2}{\cal H}_3 +
{\tanh} \frac{\beta}{2}{\cal H}_4 \right), \nonumber\\
\xi_2 &= \frac14 (- \eta_1 + \eta_2 + \eta_3 - \eta_4) =
 \frac14 \left( {- \tanh} \frac{\beta}{2}{\cal H}_1 + {\tanh} \frac{\beta}{2}{\cal H}_2 + {\tanh} \frac{\beta}{2}{\cal H}_3 -
{\tanh} \frac{\beta}{2}{\cal H}_4 \right), \nonumber\\
\xi_3 &= \frac14 (\eta_1 + \eta_2 + \eta_3 + \eta_4) =
 \frac14 \left( {\tanh} \frac{\beta}{2}{\cal H}_1 + {\tanh} \frac{\beta}{2}{\cal H}_2 + {\tanh} \frac{\beta}{2}{\cal H}_3 +
{\tanh} \frac{\beta}{2}{\cal H}_4 \right), \nonumber\\
\zeta &= \frac14 (\eta_1 - \eta_2 + \eta_3 - \eta_4) =
 \frac14 \left( {\tanh} \frac{\beta}{2}{\cal H}_1 - {\tanh} \frac{\beta}{2}{\cal H}_2 + {\tanh} \frac{\beta}{2}{\cal H}_3
- {\tanh} \frac{\beta}{2}{\cal H}_4 \right),
\end{align}
where the self-consistency fields ${\cal H}_f$ are given by the
expressions:
\begin{align}
{\cal H}_1 &=  (- \gamma_1 - \gamma_2 +
\gamma_3 + \delta), & {\cal H}_2 &=  (- \gamma_1 + \gamma_2 +
\gamma_3 - \delta), \nonumber \\
 {\cal H}_3 &=  (\gamma_1 + \gamma_2 + \gamma_3 +
\delta), &  {\cal H}_4 &=  (\gamma_1 - \gamma_2 +
\gamma_3 - \delta),
\end{align}
and
 \begin{eqnarray}
 && \gamma_1 =  \left(\frac{J_1}{2}\xi_1 +
 \mu_1E_1\right), \quad
 \gamma_2 =  \left(\frac{J_2}{2}\xi_2  +
 \mu_2E_2\right), \quad \gamma_3 =  \left(\frac{J_3}{2}\xi_3  +
 \mu_3E_3\right), \quad
 \delta =  \left(\frac{J_4}{2}\zeta + \Delta\right) .\nonumber
 \end{eqnarray}

Taking into account (\ref{eq:2.3}), we obtain
\begin{align}
J_{1} &= J_1^0 +\sum\limits_{i=1}^3\bar{\psi} _{1i}\varepsilon_i + \sum\limits_{j=4}^6\bar{\psi} _{1j}\varepsilon_j, & J_{2} &= J_2^0 +\sum\limits_{i=1}^3\bar{\psi} _{2i}\varepsilon_i + \sum\limits_{j=4}^6\bar{\psi} _{2j}\varepsilon_j,\nonumber\\
J_{3} &= J_3^0 +\sum\limits_{i=1}^3\bar{\psi} _{3i}\varepsilon_i + \sum\limits_{j=4}^6\bar{\psi} _{3j}\varepsilon_j, & J_{4} &= J_4^0 +\sum\limits_{i=1}^3\bar{\psi} _{4i}\varepsilon_i +\sum\limits_{j=4}^6\bar{\psi} _{4j}\varepsilon_j,\nonumber\\
\Delta &= \Delta^0 +\sum\limits_{i=1}^3
 \psi_{5i}\varepsilon_i + \sum\limits_{j=4}^6\psi_{5j}\varepsilon_j ,&&
\end{align}
where
\begin{align}
J_1^{0} &= - J^{0} - K_{12}^{0} + K_{13}^{0} + K_{14}^{0}, &
J_2^{0} &= J^{0} - K_{12}^{0} - K_{13}^{0} + K_{14}^{0}, \non
J_3^{0} &= J^{0} + K_{12}^{0} + K_{13}^{0} + K_{14}^{0}, &
J_4^{0} &= J ^{0}- K_{12}^{0} + K_{13}^{0} - K_{14}^{0},
 \end{align}
\begin{align}
\bar{\psi} _{1i} &= - \psi_{1i} - \psi_{2i} + \psi_{3i} + \psi_{4i}, &
\bar{\psi} _{1j} &= - \psi_{1j} - \psi_{2j} + \psi_{3j} + \psi_{4j}, \non
\bar{\psi}_{2i} &= \psi_{1i} - \psi_{2i} - \psi_{3i} + \psi_{4i}, &
\bar{\psi}_{2j} &= \psi_{1j} - \psi_{2j} - \psi_{3j} + \psi_{4j},\non
\bar{\psi}_{3i} &= \psi_{1i} +\psi_{2i} + \psi_{3i} + \psi_{4i} ,  &
\bar{\psi}_{35} &= \psi_{15} + \psi_{25} + \psi_{35} + \psi_{45} ,\non
\bar{\psi}_{4i} &= \psi_{1i} - \psi_{2i} + \psi_{3i} - \psi_{4i}, &
\bar{\psi}_{45} & = \psi_{15} - \psi_{25} + \psi_{35} - \psi_{45}. \nonumber
 \end{align}

Parameters $\xi_1$, $\xi_2$, $\xi_3$ describe the dipole
pseudospin ordering along the $a$, $b$  and $c$-axes,
respectively, and the parameter $\zeta$ is responsible for the
paraelectric phase pseudospin ordering.

Without external electric fields and mechanical strains, the
pseudospin mean values in the paraelectric phases are $\eta_1 = -
\eta_2 = \eta_3 = - \eta_4 = \eta$ and $\xi_{1p} = \xi_{2p} =
\xi_{3p} = 0$, respectively, and \be \zeta_p = {\tanh}
\frac{\beta}{2} \left( \frac{J_4}{2}\zeta_p + \Delta \right). \ee
In the ferroelectric phase at zero fields $E_i=0$ and stresses
$\sigma_j=0$, $\eta_1 = \eta_3 = \eta_{13}$, $\eta_2 = \eta_4 =
\eta_{24}$. As a result  $\xi_{1s} =0$, $\xi_{2s} =0$, and
%% 2.13
\begin{align}
\xi_{3s}  &=  \frac12 \left[ {\tanh} \frac{\beta}{2} \left(
\frac{J_3}{2} \xi_{1s}   +
\frac{J_4}{2}\zeta_s  +  \Delta \right)
  +  {\tanh} \frac{\beta}{2} \left(
\frac{J_3}{2} \xi_{1s}   -
\frac{J_4}{2}\zeta_s  -  \Delta \right) \right], \nonumber\\
\zeta_s &= \frac12 \left[ {\tanh} \frac{\beta}{2} \left(
\frac{J_3}{2} \xi_{1s}  +
\frac{J_4}{2}\zeta_s  +  \Delta \right)
  -  {\tanh} \frac{\beta}{2} \left(
\frac{J_3}{2} \xi_{1s}   -
\frac{J_4}{2}\zeta_s  -  \Delta \right) \right].
\end{align}

\section{Thermodynamic characteristics of  RHS}

To calculate the dielectric, piezoelectric, and elastic
characteristics RHS, we use the electric thermodynamical potential
per unit cell obtained in the mean field approximation
  \begin{eqnarray}
  g = \frac{G}{N} &\!\!\!=\!\!\!& v U_\textrm{seed}-
 4\frac{1}{\beta} \ln 2
  - \frac{1}{\beta}\sum\limits_{f=1}^4 \ln \cosh \frac{\beta}{2} {\cal H}_f  \non
 && + \frac{1}{2} \bigg(  J_1^0 + \sum\limits_{i=1}^3\bar{\psi_{1i}}\varepsilon_i
 +\sum\limits_{j=4}^6\bar{\psi_{1j}}\varepsilon_j\bigg)\xi_1^{2}
  +\frac{1}{2} \bigg(  J_2^0 + \sum\limits_{i=1}^3
 \bar{\psi}_{2i}\varepsilon_i+\sum\limits_{j=4}^6\bar{\psi_{2j}}\varepsilon_j\bigg)\xi_2^{2}  \nonumber\\
 && + \frac{1}{2} \bigg(  J_3^0 + \sum\limits_{i=1}^3
 \bar{\psi}_{3i}\varepsilon_i +\sum\limits_{j=4}^6\bar{\psi_{3j}}\varepsilon_j\bigg)\xi_3^{2}
  +  \frac{1}{2} \bigg(  J_4^0 + \sum\limits_{i=1}^3
 \bar{\psi}_{4i}\varepsilon_i+\sum\limits_{j=4}^6\bar{\psi_{4j}}\varepsilon_j\bigg)\zeta^{2} .
 \end{eqnarray}

From the thermodynamic equilibrium conditions
 \[
 \frac{1}{ v}\left( \frac{\partial g}{\partial
 \varepsilon_i}\right)_{E_i} =0, \qquad
 \frac{1}{ v}\left( \frac{\partial g}{\partial
 \varepsilon_j}\right)_{E_i,\sigma_i} =0, \qquad
 \frac{1}{ v}\left( \frac{\partial g}{\partial
 E_i}\right) = - P_i
 \]
 we  obtain
\begin{eqnarray}
 && 0 = c_{1i}^{E0}(T)\varepsilon_1 +
  c_{i2}^{E0}(T)\varepsilon_2 + c_{i3}^{E0}(T)\varepsilon_3  - e_{3i}^0E_3  - \frac{\bar{\psi_{1i}}}{2 v}\xi_1^{2} - \frac{\bar{\psi_{2i}}}{2
 v}\xi_2^{2} - \frac{\bar{\psi_{3i}}}{2 v}\xi_3^{2} - \frac{\bar{\psi_{4i}}}{2v}\zeta^{2} - \frac{2\psi_{5i}}{ v}\zeta,\non
 && 0 =  c_{44}^{E0}(T)\varepsilon_4  - \frac{\bar{\psi_{14}}}{2 v}\xi_1^{2} - \frac{\bar{\psi_{24}}}{2
 v}\xi_2^{2} - \frac{\bar{\psi_{34}}}{2 v}\xi_3^{2} - \frac{\bar{\psi_{44}}}{2v}\zeta^{2} - \frac{2\psi_{54}}{ v}\zeta,\non
  && 0 = c_{55}^{E0}(T)\varepsilon_5 -e_{35}^0E_3
   - \frac{\bar{\psi_{15}}}{2 v}\xi_1^{2} - \frac{\bar{\psi_{25}}}{2
 v}\xi_2^{2} - \frac{\bar{\psi_{35}}}{2 v}\xi_3^{2} - \frac{\bar{\psi_{45}}}{2v}\zeta^{2} - \frac{2\psi_{55}}{ v}\zeta,\non
  && 0 = c_{66}^{E0}(T)\varepsilon_6  - \frac{\bar{\psi_{16}}}{2 v}\xi_1^{2} - \frac{\bar{\psi_{26}}}{2
 v}\xi_2^{2} - \frac{\bar{\psi_{36}}}{2 v}\xi_3^{2} - \frac{\bar{\psi_{46}}}{2v}\zeta^{2} - \frac{2\psi_{56}}{ v}\zeta,\non
 && P_1 = e_{11}^0\varepsilon_1 + e_{12}^0\varepsilon_2 +
 e_{13}^0\varepsilon_3 + e_{15}^0\varepsilon_5 +\chi_{11}^{\varepsilon 0}E_1+ \frac{2\mu_1}{v}\xi_1, \non
 && P_2 = e_{24}^0\varepsilon_4 + e_{26}^0\varepsilon_6 + \chi_{22}^{\varepsilon 0}E_2  + \frac{2\mu_2}{v}\xi_2,\nonumber\\
 && P_3 = e_{31}^0\varepsilon_1 + e_{32}^0\varepsilon_2 +
 e_{33}^0\varepsilon_3 + e_{35}^0\varepsilon_5  +  \chi_{33}^{\varepsilon 0}E_3 + \frac{2\mu_3}{v}\xi_3.
\label{eq:3.2}
\end{eqnarray}

In the ferroelectric phase, the static isothermic dielectric
permittivities of mechanically clamped RHS along the
crystallographic axes are as follows:
%%% 3.13
\be
\label{eq:3.3}
\chi_{iis}^{T\varepsilon}(0)= \lim_{E_i\to 0} \left( \frac{\partial
P_i}{\partial E_i} \right)_{\varepsilon_j}
 = \chi_{ii}^{\varepsilon 0} +
\frac{\mu_i^2}{v}\beta F_{1is}(0).
\ee
The following notations are used
\bea && F_{11s}(0)=
 \frac{\rho_{31} - (\rho_{31}^2 -
\rho_{32}^2) \frac{\beta J_2}{4}} {1 - \rho_{31}
\left(\frac{\beta J_1}{4} + \frac{\beta J_2}{4}\right) +
(\rho_{31}^2 - \rho_{32}^2) \frac{\beta J_1}{4}\frac{\beta
J_2}{4}},
\nonumber\\
&& F_{12s}(0) =
  \frac{\rho_{31} - (\rho_{31}^2 -
\rho_{32}^2) \frac{\beta J_1}{4}} {1 - \rho_{31}
\left(\frac{\beta J_1}{4} + \frac{\beta J_2}{4}\right) +
(\rho_{31}^2 - \rho_{32}^2) \frac{\beta J_1}{4}\frac{\beta
J_2}{4}},
\nonumber\\
&& F_{13s}(0) =
  \frac{\rho_{31} - (\rho_{31}^2 -
\rho_{32}^2) \frac{\beta J_4}{4}} {1 - \rho_{31}
\left(\frac{\beta J_3}{4} + \frac{\beta J_4}{4}\right) +
(\rho_{31}^2 - \rho_{32}^2) \frac{\beta J_3}{4}\frac{\beta
J_4}{4}}, \nonumber \eea and
\[
\rho_{31} = 1 - \xi_{3s}^2 - \zeta_s^2, \qquad \rho_{32}
= 2 \xi_{3s} \zeta_s.
\]
In the paraelectric phase:
 \be \chi_{iip}^{T\varepsilon}(0) =
\chi_{ii}^{\varepsilon 0} +  \frac{\mu_i^2}{v}
\beta F_{1ip}(0), \qquad (i=1, 2, 3), \ee
where
\[
F_{1ip}(0) = \frac{1 - \zeta_p^2}{1 - (1 - \zeta_p^2)
\frac{\beta J_i}{4}}.
\]

From  relations (\ref{eq:3.3}), we get expressions for isothermic
piezoelectric coefficients  $e_{ij}$ of RHS
\begin{eqnarray}
 &&        e_{3is}^{T} =   \left( \frac{\partial P_3}{\partial
 \varepsilon_i}\right)_{E_3}   = e_{3i}^0
 + \frac{\mu_3}{v} \frac{\beta}{2}
 \left[\bar{\psi}_{3i}\xi_{3s}F_{13s}(0) -
   \left(\bar{\psi}_{4i}\zeta_s + 2 \psi_{5i}\right)
 \bar F_{13s}(0 )\right], \nonumber
 \end{eqnarray}
\begin{eqnarray}
 &&        e_{35s}^{T} =   \left( \frac{\partial P_3}{\partial
 \varepsilon_5}\right)_{E_3}   = e_{35}^0
 + \frac{\mu_3}{v} \frac{\beta}{2}
 \left[\bar{\psi}_{35}\xi_{3s}F_{13s}(0) -
   \left(\bar{\psi}_{45}\zeta_s + 2\psi_{55}\right)
 \bar F_{13s}(0 )\right]. \nonumber
 \end{eqnarray}

By differentiating the relations (\ref{eq:3.3}) with respect to the strains at
a constant polarization, we obtain the expressions for the for
piezoelectric constants
\be
 h_{3is}^{T} = \frac{e_{3is}}{\chi_{33s}^{\varepsilon}},\qquad
 h_{35s}^{T} = \frac{e_{35s}}{\chi_{33s}^{\varepsilon}}.
\ee

Now, we calculate the contributions of the pseudospin system to the
elastic constants of  RHS. From (\ref{eq:3.2}) we obtain the relations for
elastic coefficients at a constant  field:
%% 3.16
\bea
c_{ii's}^{TE} &=&  \left( \frac{\partial
\sigma_i}{\partial \varepsilon_{i'}} \right)_{E_i} = c_{ii'}^{E0}(T) -
 \frac{{\beta}\bar{\psi_{3i}}\bar{\psi_{3i'}}}{4 v} \xi_{3s}^2 F_{13s}(0)-
\frac{{\beta}\left(\bar{\psi_{4i}}\zeta_s+2\psi_{5i}\right)
\left(\bar{\psi_{4i'}}\zeta_s+2\psi_{5i'}\right)}{4 v}  F_{14s}(0) \nonumber \\
&&\phantom{\left( \frac{\partial
\sigma_i}{\partial \varepsilon_{i'}} \right)_{E_i}=}+\frac{\beta\left(\bar{\psi_{3i}}\bar{\psi_{4i'}}+\bar{\psi_{3i'}}\bar{\psi_{4i}}\right)}{4\upsilon}\xi_{3s}\zeta_s\bar F_{13s}(0) + \frac{\beta\left(\bar{\psi_{4i}}\psi_{5i'}+\bar{\psi_{4i'}}\psi_{5i}\right)}{2\upsilon}\xi_{3s}\bar F_{13s}(0),\\ \nonumber
c_{ii'p}^{TE} &=&  c_{ii'}^{E0}(T) -
\frac{{\beta}\left(\bar{\psi_{4i}}\zeta_{p}+2\psi_{5i}\right)
\left(\bar{\psi_{4i'}}\zeta_{p}+2\psi_{5i'}\right)}{ 4 v}  F_{14p}(0 ),\\ \nonumber
 c_{jjs}^{TE} &=&  \left( \frac{\partial
\sigma_j}{\partial \varepsilon_j} \right)_{E_i} = c_{jj}^{E0}(T) - \frac{{\beta}\bar{\psi_{3j}^2}}{4 v} \xi_{3s}^2 F_{13s}(0)-
\frac{{\beta}\left(\bar{\psi_{4j}}\zeta_s+2\psi_{5j}\right)^{2}}{4 v}  F_{14s}(0)\nonumber\\
&& \phantom{\left( \frac{\partial
\sigma_j}{\partial \varepsilon_j} \right)_{E_i} =} + \frac{\beta\bar{\psi_{3j}}\bar{\psi_{4j}}}{2\upsilon}\xi_{3s}\zeta_s\bar F_{13s}(0)+ \frac{\beta\bar{\psi_{3j}}\psi_{5j}}{\upsilon}\xi_{3s}\bar F_{13s}(0),\\
\nonumber
 c_{jjp}^{TE} &=&  c_{jj}^{E0}(T) -
\frac{{\beta}\left(\bar{\psi_{4j}}\zeta_s+2\psi_{5j}\right)^{2}}{ 4 v}  F_{14p}(0 ),
 \eea
 where
 \begin{eqnarray}
 &&         F_{14s}(0) = \frac{ \rho_{31} - ( \rho_{31}^2 -  \rho_{32}^2)\frac{\beta J_{3}}{4}}
 {1 - \rho_{31} (\frac{\beta J_{3}}{4} + \frac{\beta J_{4}}{4}) + (\rho_{31}^2 - \rho_{32}^2)\frac{\beta J_{3}}{4}\frac{\beta J_{4}}{4} },
\qquad  F_{14p}(0) = \frac{1- \zeta_p^2}{1- (1- \zeta_p^2)\frac{\beta
 J_{4}}{4}}. \nonumber
 \end{eqnarray}

From (\ref{eq:3.2}), (\ref{eq:3.3}) we get isothermic coefficients of
piezoelectric strain $d_{1i} = ( {\partial P_1}/{\partial
\sigma_i} )_{E_1}$, \linebreak $d_{ij} = ( {\partial P_i}/{\partial \sigma_j})_{E_i}$
in the following form:
\begin{eqnarray}
 && d_{3i}^{T} = \sum\limits_{k=1}^3 s_{ik}^{TE} e_{3k}^{T} + s_{i5}^{TE}
 e_{35}^{T}, \qquad  d_{35}^{T} = s_{15}^{TE}e_{31}^{T} + s_{25}^{TE}e_{32}^{T} + s_{35}^{TE}e_{33}^{T} + s_{55}^{TE}e_{35}^{T},
  \nonumber
\end{eqnarray}
where  $s_{ik}^E = ( {\partial \varepsilon_i}/{\partial \sigma_k})_{E_1}$,
$s_{jj}^E = ( {\partial \varepsilon_j}/{\partial \sigma_j} )_{E_i}$
are the compliances at the constant field.

Using relations (\ref{eq:3.3}) one can obtain the expression for the static dielectric permittivity of a free RHS crystal
 \begin{eqnarray}
 && \chi_{33}^{T\sigma} = \left( \frac{\partial P_3}{\partial E_3}
 \right)_{\sigma_6} = \chi_{23}^{T\varepsilon} +  e_{31}^{T}d_{31}^{T} +
 e_{32}^{T}d_{32}^{T}+ e_{33}^{T}d_{33}^{T} + e_{35}^{T}d_{35}^{T}. \nonumber
 \end{eqnarray}

Molar entropy of RHS caused by its pseudospin subsystem is as
follows:
\begin{equation}\label{entrop}
S=-R\left(\frac{\partial g}{\partial T}\right) = R\bigg[4\ln2+\sum\limits_{f=1}^4 \ln
{\cosh} \frac{\beta}{2} {\cal H}_f - 2\gamma_1\xi_1-2\gamma_2\xi_2
-2\gamma_3\xi_3-2\delta\varsigma\bigg],
\end{equation}
where $R$ is the universal gas constant. Molar heat capacity at
a constant pressure is calculated by differentiating the entropy
(\ref{entrop})
\begin{equation}\label{DC}
\Delta C^{\sigma} = T\left(\frac{\partial S}{\partial T}\right)_{\sigma}.
\end{equation}

\section{Relaxation dynamics of RHS crystal}

This section describes the dynamic phenomena in RHS at the
application of  electrical field  $E_1^{\ast}$ to a crystal. While
calculating the dynamic characteristics, we use the kinetic
equation \cite{320x,321x} based on the Zubarev nonequilibrium
statistical operator method \cite{38x}.

The kinetic equation for the mean values of pseudospin operator is
as follows:
\be
\frac{\rd}{\rd t} \langle \hat p_m \rangle = - \sum\limits_{qf} \sum\limits_{\mu\alpha} \left[ Q_{qf\mu\alpha}^{-}(\hat p_m) + {\tanh} \frac{\beta \Omega_{\mu}^{\alpha}}{2} Q_{qf\mu\alpha}^{+}(\hat p_m)\right] K_{\mu}^{\alpha}\label{kin.riv},
\ee
where
\bea
&& Q_{qf\mu\alpha}^{\mp}(\hat p_m) = \langle \left[ \bigl[ \hat p_m, \sigma_{qf}^{-\alpha} \bigl( \Omega_{\mu}^{\alpha'} \bigr) \bigr], \sigma_{qf}^{\alpha} \bigl( \Omega_{\mu}^{\alpha} \bigr) \right]^{\mp} \rangle_q, \label{dP}\\
&& K_{\mu}^{\alpha} = \int\limits_0^{\infty} \rd t \re^{-\varepsilon
t}\cos \Omega_{\mu}^{\alpha}t{\rm Re} \langle \bar u(t)\bar u^{+}
\rangle_q, \qquad \alpha=0,\pm 1,
\eea
while $\langle \bar
u^{\alpha}(t)\bar u^{\alpha'} \rangle_q$ is correlation function
of the thermostat; $\sigma_{qf}^{\alpha} \bigl(
\Omega_{\mu}^{\alpha} \bigr)$ is a Fourier component of the
operator $\sigma_{qf}^{\alpha}(t)$; $\Omega_{\mu}^{\alpha}$ are
the eigenfrequencies of the Hamiltonian of the quasispin model
(\ref{eq:2.7}); $\sigma_{qf}^0 = \sigma_{qf}$, $\sigma_{qf}^{\pm} =
\sigma_{qf}^x \pm i\sigma_{qf}^y$.

Taking into account (\ref{RReq2.5}), operators $\hat{p}_m$ have such a form:
\be \hat{P}_m =\frac{\sigma_{qf}}{2}. \ee

Using the evolution law of the  quasispin operators
$S_{qf}^{\alpha}$ ($\alpha=0\pm$) and their permutation relations,
we calculate the commutators occurring in (\ref{dP}) as well as
the expression for $Q_{qf\mu\alpha}^{\mp}(\hat p_m)$. The kinetic
equation (\ref{kin.riv}) can be rewritten as follows:
 \be -\frac{\rd}{\rd t}\eta_f = 2K_f\eta_f
-2K_f \tanh\frac{\beta H_f}{2}, \label{deta} \ee
where
\[  K_f = \int\limits_0^{\infty}\rd t \re^{-\varepsilon
t}\cos(H_ft)\textrm{Re}[\langle \bar{u}^-(t)\bar{u}^+\rangle_q + \langle
\bar{u}^+(t)\bar{u}^-\rangle_q]. \]
Note that at $K_f=\frac{1}{2\alpha}$, the obtained kinetic equation
(\ref{deta}) agrees with the equation found within the stochastic
Glauber model \cite{31x}. Using the variables $\xi_1$, $\xi_2$,
$\xi_3$, $\zeta$ in equations (\ref{deta}), we obtain
 \bea
&& - \alpha \frac{\rd}{\rd t} \xi_1 = \xi_1 - \frac14(-L_1-L_2+L_3+L_4), \nonumber\\
&& - \alpha \frac{\rd}{\rd t} \xi_2 = \xi_2 - \frac14(-L_1-L_2-L_3-L_4), \nonumber\\
&& - \alpha \frac{\rd}{\rd t} \xi_3 = \xi_3 - \frac14(L_1+L_2+L_3+L_4), \nonumber\\
&& - \alpha \frac{\rd}{\rd t} \zeta = \zeta - \frac14(L_1-L_2+L_3-L_4), \label{3.1}
 \eea
where the following notations are used:
%%% 3.2
\begin{align}   L_{1} &= {\tanh} \frac12 (-\gamma_1 - \gamma_2
+ \gamma_3 + \delta), &  L_{2} &= {\tanh} \frac12 (-\gamma_1 +
\gamma_2 +
\gamma_3 - \delta),   \nonumber\\
L_{3} &= {\tanh} \frac12 (\gamma_1 + \gamma_2 +
\gamma_3 + \delta), & L_{4} &= {\tanh} \frac12 (\gamma_1 -
\gamma_2 + \gamma_3 - \delta). \label{3.2}
\end{align}

The  dynamic properties RHS are explored using the system of
equations (\ref{3.1}) and at small deviations from the equilibrium. We
separate these equations into the static and dynamic parts. The
distribution functions are presented as sums of two components:
the equilibrium functions and their deviations from the
equilibrium values (fluctuations)
\bea && \xi_1 = \tilde
\xi_1 + \xi_{it}, \qquad (i=1,2,3), \qquad
\zeta = \tilde \zeta + \zeta_t.
\eea
Also $E_{it}=E_{i0}\re^{\ri\omega t}$.

As a result, we obtain the  following  system of equations for
the fluctuation parts: \be - \frac{\rd}{\rd t} \left( \begin{array}{c}
            \xi_{1ts}(1) \\
            \xi_{2ts}(1)
            \end{array}
            \right) =
        \left( \begin{array}{cc}
            a_{11} & a_{12} \\
            a_{21} & a_{22}
            \end{array}
            \right)
        \left( \begin{array}{c}
            \xi_{1ts}(1) \\
            \xi_{2ts}(1)
            \end{array}
            \right) - \frac{\beta \mu_1 E_{1t}}{2}
        \left( \begin{array}{c}
            a_1 \\
            a_2
            \end{array}
            \right), \label{3.8}
\ee
%%% 3.9
\be - \frac{\rd}{\rd t} \left( \begin{array}{c}
            \xi_{2ts}(2) \\
            \xi_{1ts}(2)
            \end{array}
            \right) =
        \left( \begin{array}{cc}
            b_{11} & b_{12} \\
            b_{21} & b_{22}
            \end{array}
            \right)
        \left( \begin{array}{c}
            \xi_{2ts}(2) \\
            \xi_{1ts}(2)
            \end{array}
            \right) - \frac{\beta \mu_2 E_{2t}}{2}
        \left( \begin{array}{c}
            b_1 \\
            b_2
            \end{array}
            \right), \label{3.9}
\ee
%%% 3.10
\be - \frac{\rd}{\rd t} \left( \begin{array}{c}
            \xi_{3ts}(3) \\
            \zeta_{st}(3)
            \end{array}
            \right) =
        \left( \begin{array}{cc}
            c_{11} & c_{12} \\
            c_{21} & c_{22}
            \end{array}
            \right)
        \left( \begin{array}{c}
            \xi_{3ts}(3) \\
            \zeta_{st}(3)
            \end{array}
            \right) - \frac{\beta \mu_3 E_{3t}}{2}
        \left( \begin{array}{c}
            c_1 \\
            c_2
            \end{array}
            \right). \label{3.10}
\ee Solving the systems (\ref{3.8})--(\ref{3.10}), we find the
dynamic permittivities of the clamped RHS crystal
\bea &&
\varepsilon_{iis}^{\varepsilon}(\omega) =
\varepsilon_{iis}^{\varepsilon 0} + \sum\limits_{j=1}^2
\frac{4\pi\chi_{jis}}{1+ (\omega \tau_{jis})^2} + i
\sum\limits_{j=1}^2 \frac{4\pi\omega\tau_{jis}\chi_{jis}}{1+
(\omega \tau_{jis})^2}. \label{3.11}
\eea
In  (\ref{3.11})
\bea
          && \chi_{1is} = \frac{\mu_i^2}{v} \beta
\frac{\tau_{1is}\tau_{2is}} {\tau_{1is} - \tau_{2is}} \bigg[
-m^{(1)} (i) +\tau_{1is}  m^{(0)} (i) \bigg],
\qquad
\chi_{2is} =
\frac{\mu_i^2}{v} \beta \frac{\tau_{1is}\tau_{2is}} {\tau_{1is} -
\tau_{2is}} \bigg[ m^{(1)} (i) -\tau_{2is}  m^{(0)} (i) \bigg],
\label{3.12} \nonumber
\eea
\be
\tau_{{1\atop 2}is}^{-1} = \frac12
\bigg[m_1(i)\mp \sqrt{m_1^2(i) - 4m_0(i)}\bigg]. \label{3.13}
\ee
In (\ref{3.13}), we use the following notations:
\bea
&&  m_1(1) = m_1(2) =    \frac{1}{\alpha} \left( 1 -   \rho_{31} \frac{\beta J_1}{4} \right) + \frac{1}{\alpha} \left(   1 - \rho_{31} \frac{\beta J_2}{4}   \right), \nonumber\\
&&  m_0(1) = m_0(2)  = \frac{1}{\alpha^2} \left[  1 -   \rho_{31} \left( \frac{\beta J_1}{4} +   \frac{\beta J_2}{4} \right) +   \Bigl( \rho_{31}^{2}   -    \rho_{32}^2 \Bigr) \frac{\beta J_1}{4} \frac{\beta J_2}{4} \right], \nonumber\\
&&  m^{(0)}(1) = \frac{1}{\alpha^2} \left[ \rho_{31} - \left(  \rho_{31}^2 -  \rho_{32}^2 \right) \frac{\beta J_2}{4} \right], \qquad
m^{(0)}(2) = \frac{1}{\alpha^2} \left[ \rho_{31} - \left( \rho_{31}^2 -  \rho_{32}^2
\right) \frac{\beta J_1}{4} \right]. \nonumber\\
&&  m_1(3) =  \frac{1}{\alpha}  \left( 1 - \rho_{31} \frac{\beta J_3}{4} \right) + \frac{1}{\alpha} \left( 1 - \rho_{31} \frac{\beta J_4}{4} \right), \nonumber\\
&&  m_0(3) = \frac{1}{\alpha^2} \left[ 1 - \rho_{31} \left( \frac{\beta J_3}{4} + \frac{\beta J_4}{4} \right) + \left( \rho_{31}^{2} -  \rho_{32}^2 \right) \frac{\beta J_3}{4} \frac{\beta J_4}{4} \right], \nonumber\\
&&  m^{(1)}(3) =  \frac{1}{\alpha} \rho_{31}, \qquad
m^{(0)}(3) = \frac{1}{\alpha^2} \left[  \rho_{31} - \left(
\rho_{31}^{2} -
 \rho_{32}^2 \right)  \frac{\beta J_4}{4}  \right]. \nonumber
 \eea

\section{Comparison of numerical results with   experimental data}

To compare the  temperature and field dependences of the above derived
 dielectric, piezoelectric, elastic, and thermal
characteristics of RHS, we need to set the values of the following
parameters: the interaction potentials $J^{0}$, $K^{0}_{12}$,
$K^{0}_{13}$, $K^{0}_{14}$ and, accordingly, $J_{1}$, $J_{2}$,
$J_{3}$, $J_{4}$; the parameter $\Delta$, which characterizes an
asymmetry of populations of the two equilibrium positions of a
dipole; the deformation potentials $\psi_{ij}$; effective dipole
moments $\mu_i$; seed dielectric permittivities
$\chi_{ii}^{\varepsilon 0}$; piezoelectric coefficients
$e_{ij}^0$; elastic constants $c_{ii'}^{E0}$ and $c_{jj}^{E0}$,
and the parameter $\alpha$ that defines the time scale of the
relaxation processes.

To find the optimum values of the theory parameters it is
necessary to use the dependence of temperature $T_\textrm{c}$ on
hydrostatic pressure. Unfortunately,  different sources give
different values for $T_\textrm{c}(0)$, varying from 258~K to 265.1~K. We
shall use $T_\textrm{c}(0) = 265$~K \cite{gri}.

In the fitting procedure, we  use the experimentally obtained
values for the  temperature dependences of the following physical
characteristics of RHS: $P_s(T)$ \cite{kaj},
$\varepsilon_{11}^\sigma(0)$, $\varepsilon_{22}^\sigma(0)$
\cite{zaj}, $\varepsilon_{33}(\omega)$ \cite{gri}, as well as the
dependence $T_\textrm{c}(p)$ \cite{ges} of the transition temperature on
 hydrostatic pressure. In the case of deuterted RDS crystal, we exploit $P_s(T)$
\cite{ich}, $\varepsilon_{33}^\sigma(0)$ \cite{ich},  $T_\textrm{c}(p)$
\cite{ges}.

In order to find the values of the parameters
$J^{0}+K^{0}_{13}$, $K^{0}_{12}+K^{0}_{14}$, $\Delta$,
we found the point at the phase diagram $(a,b)$, where
\[a=\frac{(J^{0}+K^{0}_{13})-(K^{0}_{12}+K^{0}_{14})}
{(J^{0}+K^{0}_{13})+(K^{0}_{12}+K^{0}_{14})},
\qquad
b=\frac{\Delta}{(J^{0}+K^{0}_{13})+(K^{0}_{12}+K^{0}_{14})},
\]
at which the system undergoes a single
 second order phase transition at
 $T_\textrm{c}(0) = 265$~K.
 The values of the
  effective dipole moments $\mu_3$
were determined by fitting the calculated spontaneous polarization to experiment.

The values of
 $J^{0}$, $K^{0}_{12}$, $K^{0}_{13}$, $K^{0}_{14}$, $\mu_1$, and $\mu_2$
are determined by fitting the calculated  $\varepsilon_{11}$ and $\varepsilon_{22}$ to the
experimental data given in \cite{zaj}.

The fitting procedure for the models of this class is described in detail in  \cite{zac}, where the thermodynamic characteristics of Rochelle salt are explored.

The strains should be taken into account in order to calculate the dielectric permittivity of a mechanically free crystal, piezoelectric coefficients, and
elastic constants. Therefore, to determine the values of the  deformation potentials $\psi_{ij}$, we analyzed their effect on theoretical values of physical characteristics of the crystal.
Thus, it has been obtained that an increase of the transition temperature with hydrostatic pressure
 \cite{ges} can be described using the values of  $\psi_{ij}$ presented here. It should be stressed that
 when the piezoelectric coefficients are measured experimentally, the values of  $\psi_{ij}$ can be determined with a
 greater accuracy.

The parameter $\alpha_H$ is determined from the condition that
the theoretically calculated frequency dependences of
$\varepsilon_{33}(\omega)$  agree with the experiment. We also assume
that the parameter $\alpha_H$ is a weak function of temperature:
\[
\alpha_H = [P_H + R_H(\Delta T)]\times 10^{-14}, \qquad \Delta T = T - T_\textrm{c}.
\]
The unit cell volume of RHS  is $v = 0.842\cdot 10^{-21}$~cm$^3$.

The obtained sets of optimal parameters are given in table~\ref{tab1}.

\renewcommand{\arraystretch}{1.0}
\renewcommand{\tabcolsep}{2.2pt}
\begin{table}[!h]
\caption{The  optimal sets of the theory parameters for
Rb(H$_{1-x}$D$_x)$SO$_4$.}\label{tab1}
\begin{center}
\begin{tabular}{|c|c|c|c|c|c|c|c|c|c|c|c|}
\hline\hline $x$ &$J^{0}/k_\textrm{B}$& $K_{12}^{0}/k_\textrm{B}$& $K_{13}^{0}/k_\textrm{B}$& $K_{14}^{0}/k_\textrm{B}$&$\Delta / k_\textrm{B}$ & $\mu_{1}$, $10^{-18}$& $\mu_{2}$, $10^{-18}$& $\mu_{3}$, $10^{-18}$& $\chi_{11}^{\varepsilon 0}$&$\chi_{22}^{\varepsilon 0}$&$\chi_{33}^{\varepsilon 0}$\\
 &K& K& K& K& K &  esu$\cdot$cm & esu$\cdot$cm & esu$\cdot$cm &&& \\
\hline 0.0 &394& 190& 372& 433.7& 244& 3.18  & 3.65& 0.81&0.02&0.02&0.159  \\
\hline 0.7& 380& 189& 345& 430.8& 245&   & & 0.90&0.02&0.02&0.159  \\
\hline 1.0& 378& 198.8& 338.4& 429& 245.4&   & & 1.00&0.02&0.02&0.159  \\
\hline\hline
\end{tabular}
\\[2ex]
\begin{tabular}{|c|c|c|c|c|c|c|}
\hline\hline
  $x$ &$P_{3s}$ & $R_{3s}$ &$P_{3p}$ &$P_{3p}$ & $P_{1,2}$ & $R_{1,2}$ \\
   &(s) & (s/K)  & (s)  &  (s/K)&(s) & (s/K)  \\
\hline 0.0&12.5 & --0.0521 & 12.5 & --0.091& 10.5& --0.001 \\
\hline 0.7&10.7 & --0.0510 & 10.7 & --0.051& &  \\
\hline 1.0&9.8 & --0.0510 & 9.8 & --0.0501& &  \\
\hline\hline
\end{tabular}
\end{center}
\end{table}

The deformation potentials are taken to be
${\bar{\psi} _{1i}}/{k_\textrm{B}}= {\bar{\psi} _{1j}}/{k_\textrm{B}}= {\bar{\psi} _{2i}}/{k_\textrm{B}}= {\bar{\psi} _{2j}}/{k_\textrm{B}}=900$~K,
${\bar{\psi} _{31}}/{k_\textrm{B}}=-4950$~K,
${\bar{\psi} _{32}}/{k_\textrm{B}}= {\bar{\psi} _{33}}/{k_\textrm{B}}= {\bar{\psi} _{35}}/{k_\textrm{B}}= -4500$~K,
${\bar{\psi} _{41}}/{k_\textrm{B}}=1080$~K,
${\bar{\psi} _{42}}/{k_\textrm{B}}= {\bar{\psi} _{43}}/{k_\textrm{B}}= {\bar{\psi} _{45}}/{k_\textrm{B}}= 900$~K,
${\bar{\psi} _{51}}/{k_\textrm{B}}= {\bar{\psi} _{52}}/{k_\textrm{B}}
= {\bar{\psi} _{53}}/{k_\textrm{B}}= {\bar{\psi} _{55}}/{k_\textrm{B}}= 200$~K.

The ``seed'' constants for RHS are $e_{31}^0 = e_{32}^0 = e_{33}^0 = e_{35}^0 = -1\times 10^4$~{esu}/{cm$^2$}, $c_{11}^{0E} = 32.0 \times 10^{10}$~{dyn}/{cm$^2$},
$c_{12}^{E0} = 17.0 \times 10^{10}$~{dyn}/{cm$^2$},
$c_{13}^{E0} = 8.7 \times 10^{10}$~{dyn}/{cm$^2$},
$c_{22}^{E0} = 38 \times 10^{10}$~{dyn}/{cm$^2$},
$c_{23}^{E0} = 6.5 \times 10^{10}$~{dyn}/{cm$^2$},
$c_{33}^{E0} = 37.4 \times 10^{10}$~{dyn}/{cm$^2$},
$c_{44}^{E0} = 4.9 \times 10^{10}$~{dyn}/{cm$^2$},
$c_{55}^{E0} = 5.3 \times 10^{10}$~{dyn}/{cm$^2$},
$c_{66}^{E0} = 12.8 \times 10^{10}$~{dyn}/{cm$^2$};
$k_{11}=-0.032\times 10^{10}$~{dyn}/(cm$^2$K),
$k_{12}=-0.040\times 10^{10}$~{dyn}/(cm$^2$K),
$k_{13}=-0.015\times 10^{10}$~{dyn}/(cm$^2$K),
$k_{23}=-0.010\times 10^{10}$~{dyn}/(cm$^2$K)
$k_{33}=-0.032\times 10^{10}$~{dyn}/(cm$^2$K),
$k_{22}=k_{44}=k_{55}$=0.0. We use the same values for RDS as
well.

\begin{figure}[!t]
\begin{center}
\includegraphics[width=0.44\textwidth]{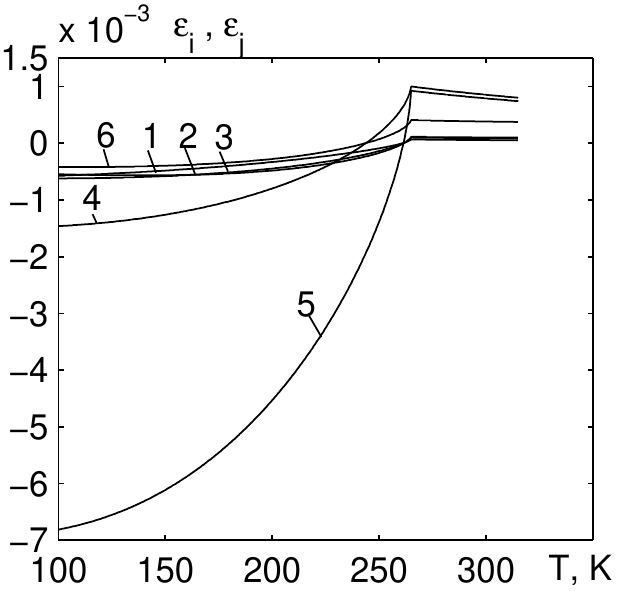}
\hspace{5mm}
\includegraphics[width=0.42\textwidth]{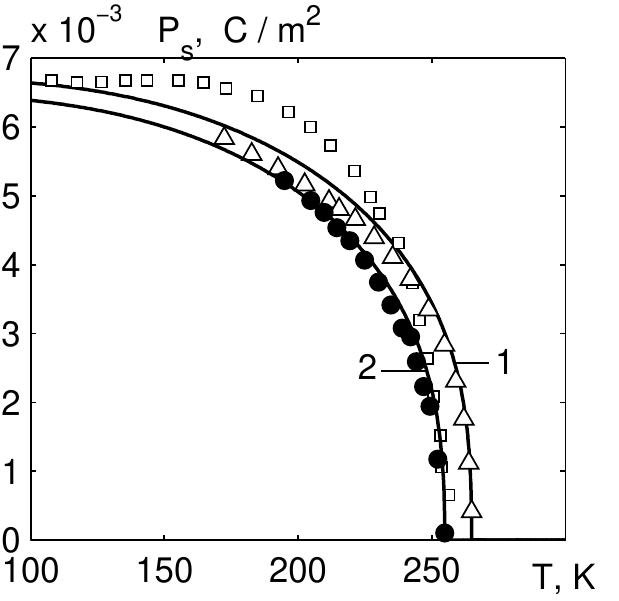}
\parbox[t]{0.49\textwidth}{%
\vspace{-3mm}
\caption[]{The temperature dependences of the strains
$\varepsilon_i$ and $\varepsilon_j$  of RHS:
 $\varepsilon_1$~--- 1, $\varepsilon_2$~--- 2, $\varepsilon_3$~--- 3,$\varepsilon_4$~--- 4, $\varepsilon_5$~--- 5 i $\varepsilon_6$~--- 6.} \label{epsi,epsj}
}
\hfill%
\parbox[t]{0.49\textwidth}{%
\vspace{-3mm}
\caption[]{The temperature dependences of spontaneous polarization
of RHS~--- 1, $\vartriangle$ \cite{kaj}, $\square$ \cite{pep} and
RDS~--- 2, $\bullet$ \cite{ich}.} \label{Psxd}
}
\end{center}
\end{figure}

Now we discuss the obtained results. In figure~\ref{epsi,epsj},
the temperature dependences of the strains $\varepsilon_i$ and $\varepsilon_j$ are presented. In the ferroelectric phase,
$\varepsilon_i$ slightly increases with temperature, while the temperature variation of   $\varepsilon_4$, $\varepsilon_6$, and
especially $\varepsilon_5$  is much stronger. In the paraelectric phase, all these strains
weakly increase with temperature.
The temperature dependences of spontaneous polarization $P_s$ of
RHS and RDS along with the  experimentally obtained values
\cite{kaj,pep,ich} are shown in figure~\ref{Psxd}. A good
description of experimental data of \cite{kaj} and \cite{ich} is
reached. When the deuteration level $x$ increases, the
polarization decreases.

\begin{figure}[!h]
\vspace{-3mm}
\begin{center}
\hspace{-3mm}
\includegraphics[width=0.44\textwidth]{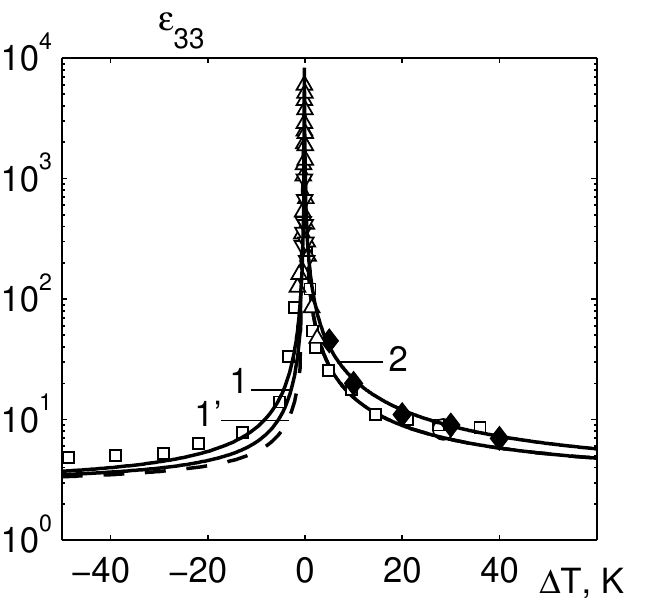}
\hspace{5mm}
\includegraphics[width=0.41\textwidth]{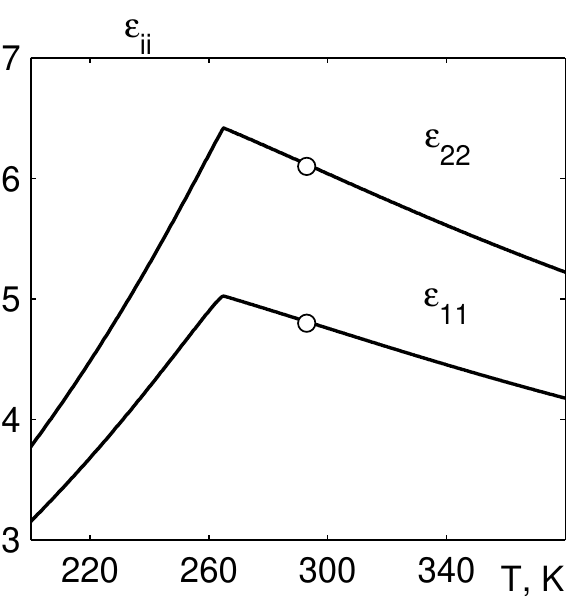}
\parbox[t]{0.49\textwidth}{%
\vspace{-3mm}
\caption[]{The temperature dependences of static dielectric
permittivities of  RHS, 1~--- $\varepsilon^{\sigma}_{33}$, 1'~---
$\varepsilon^{\varepsilon}_{33}$,  $\triangledown$ \cite{fle},
$\vartriangle$ \cite{kaj}, $\square$ \cite{pep}, $\circ$ \cite{zaj}
and RDS~--- 2, $\blacklozenge$ \cite{ich}.} \label{eps33}
}
\hfill
\parbox[t]{0.49\textwidth}{%
\vspace{-3mm}
\caption[]{The temperature dependences of the transverse
permittivities $\varepsilon_{11}$ and $\varepsilon_{22}$ of RHS.
$\circ$ \cite{zaj}.} \label{epsii}
}
\end{center}
\end{figure}

 Figure~\ref{eps33} shows the calculated temperature dependences of
static dielectric permittivities of the mechanically clamped
$\varepsilon_{33}^\varepsilon(0,T)$ and free
$\varepsilon_{33}^\sigma(0,T)$ RHS crystal along with the
experimental data \cite{fle,kaj,pep,zaj}.

\begin{figure}[!t]
\vspace{-3mm}
\begin{center}
\includegraphics[scale=0.9]{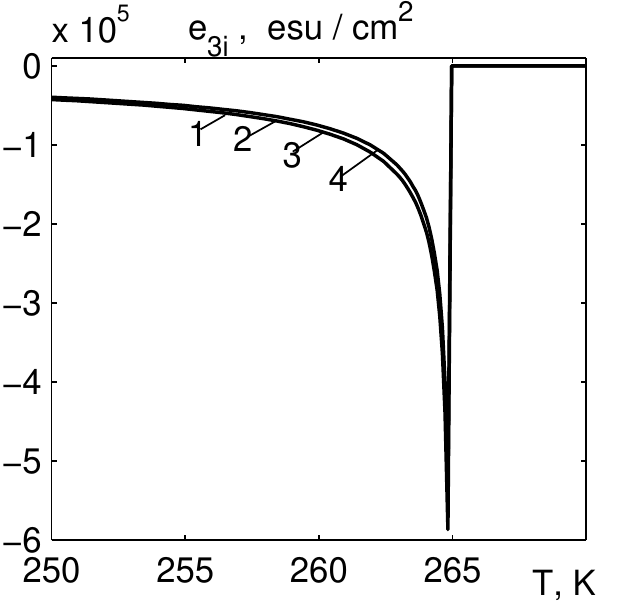}
\hspace{5mm}
\includegraphics[scale=0.9]{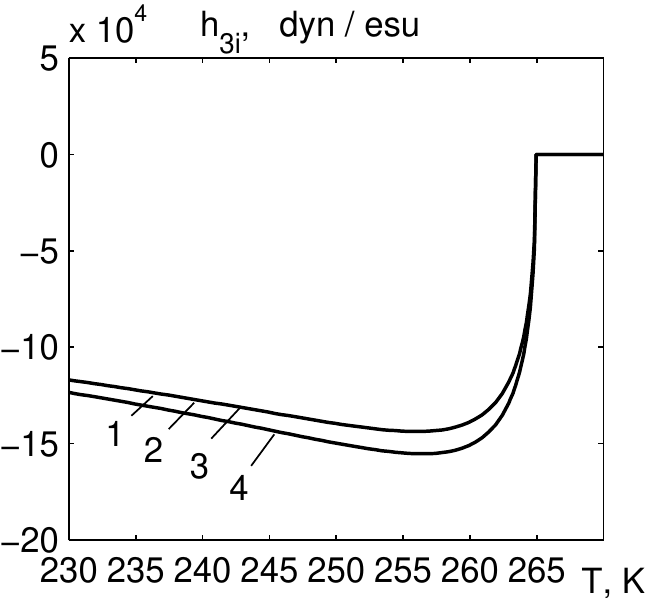}
\end{center}
\vspace{-3mm}
\caption[]{The temperature dependences of piezoelectric
coefficients $e_{31}$~---1, $e_{32}$~---2, $e_{33}$~---3, $e_{35}$~---4
and constants  $h_{31}$~---1, $h_{32}$~---2, $h_{33}$~---3, $h_{35}$~---4
of RHS.} \label{e3i}
\end{figure}

The permittivity $\varepsilon_{33}^\sigma(0,T)$  is larger than
$\varepsilon_{33}^\varepsilon(0,T)$. An increase of deuteron
concentration increases the permittivity
$\varepsilon_{33}^\varepsilon(0,T)$ at all temperatures.
As shown in figure~\ref{eps33}, the theoretical results $\varepsilon_{33}^\sigma(0,T)$ are in a good quantitative agreement with
experimental data of \cite{fle,kaj,pep,zaj}. At temperature $T = T_\textrm{c}$,  the value of the permittivity $\varepsilon_{33}^\sigma(0,T)$
is very large, which is  typical of the second order phase transitions.

Figure~\ref{epsii} illustrates the temperature  dependences of the
transverse permittivities $\varepsilon_{11}$ and
$\varepsilon_{22}$ of a RHS crystal. They are significantly
smaller than the longitudinal permittivity.

The temperature dependences of piezoelectric  coefficients
$e_{3i}$ and $e_{55}$ and constants $h_{3i}$ and $h_{35}$ are
given in figure~\ref{e3i}. In the paraelectric phase, these
coefficients are equal to zero, whereas in the ferroelectric phase,
$e_{3i}$ and $e_{55}$ values have a deep minimum at approaching
$T_\textrm{c}$, and $h_{31}$, $h_{35}$ constants change insignificantly.

\begin{figure}[!h]
\begin{center}
\hspace{-5mm}
\includegraphics[scale=0.8]{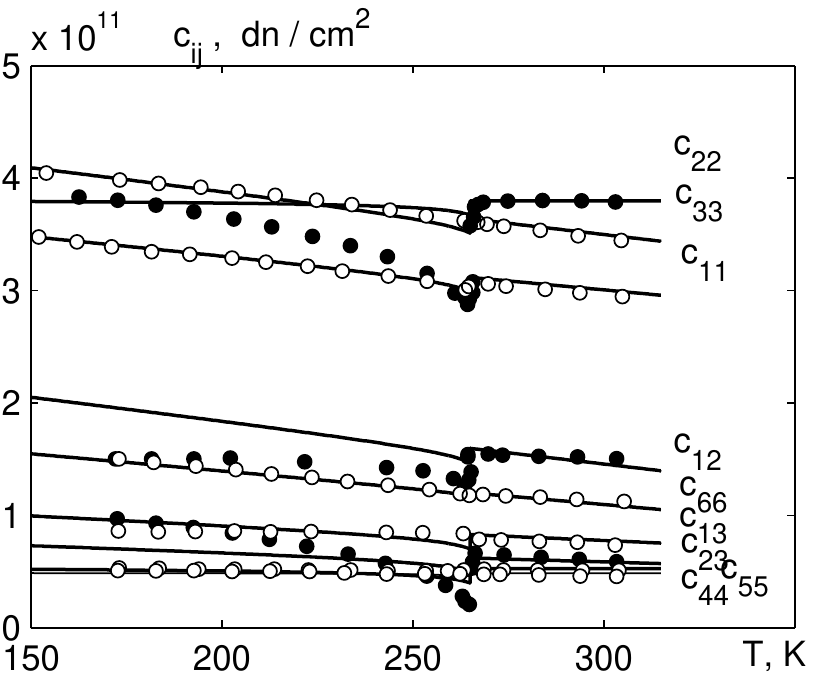}
\hspace{5mm}
\includegraphics[scale=0.9]{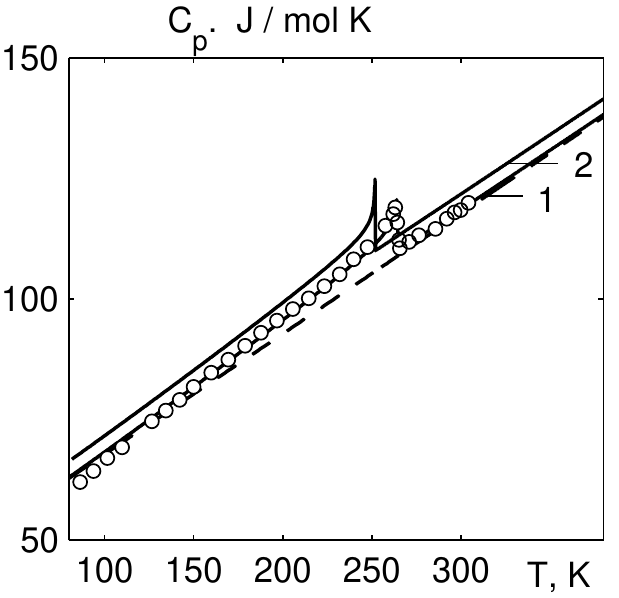}
\parbox[t]{0.5\textwidth}{%
\vspace{-2mm}
\caption[]{The temperature dependences of the elastic constants
$c_{ij}^E$  of RHS, $\circ$~--- \cite{zaj}, lines: the theory.}\label{cij}
}
\hfill
\parbox[t]{0.49\textwidth}{%
\vspace{-2mm}
\caption[]{The temperature dependences of heat capacity of RHS~--- 1, $\bullet$ \cite{ale} and  RDS~--- 2.}
\label{cp}
}
\end{center}
\end{figure}

In figure~\ref{cij} we show the temperature dependences of the
elastic constants $c_{ij}^E$. A good description of experimental
data  \cite{zaj} is obtained, except for the  temperature
dependence of elastic constants  $c_{12}^E$, $c_{22}^E$, and
$c_{23}^E$ in the ferroelectric phase, for which the experimental
measurements predict deep minima near~$T_\textrm{c}$.

\begin{figure}[!h]
\vspace{-3mm}
\begin{center}
\includegraphics[scale=0.9]{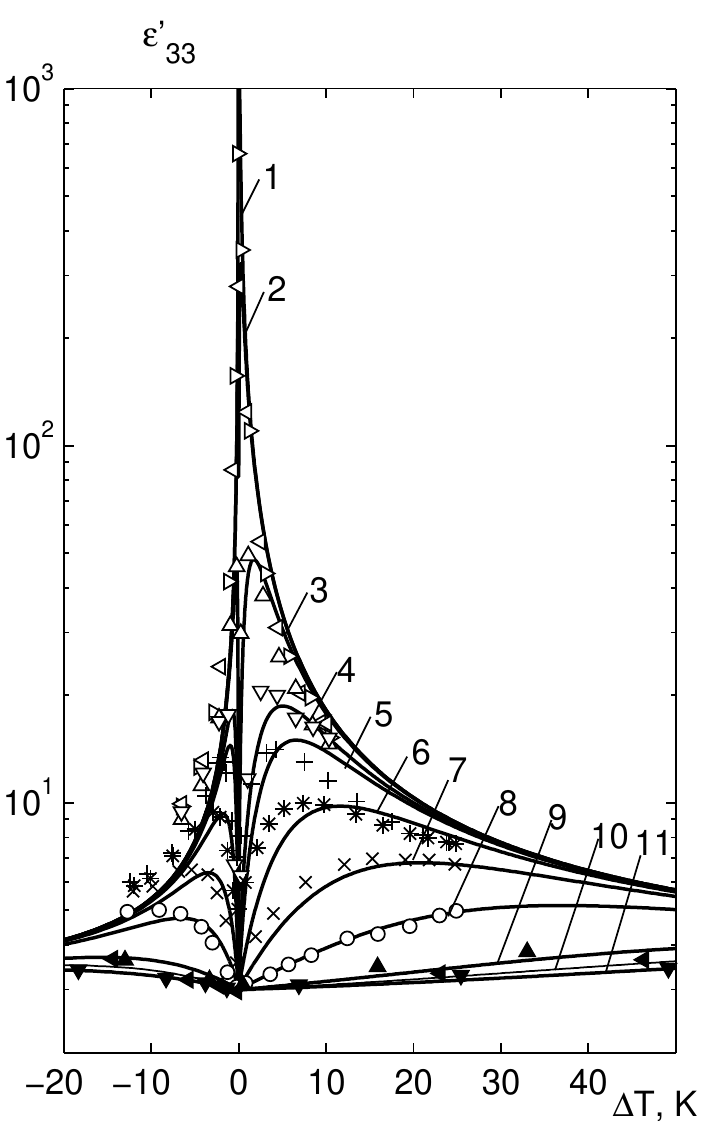}
\hspace{0.5cm}
\includegraphics[scale=0.9]{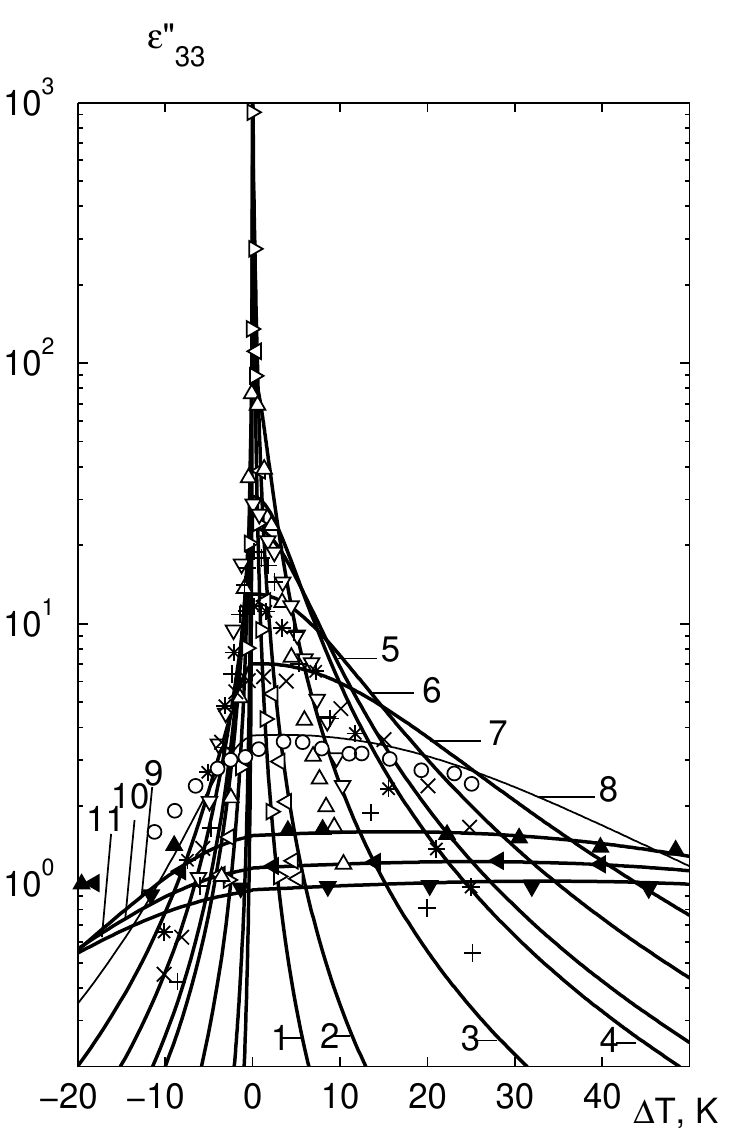}
\vspace{-2mm}
\caption[]{The temperature dependences of the real
$\varepsilon_{33}^{\varepsilon'}(\nu,T)$ and imaginary
$\varepsilon_{33}^{\varepsilon''}(\nu,T)$ parts of the dynamic
dielectric permittivity of RHS at different frequencies $\nu$
(GHz): 0.150~--- 1, $\triangleright$ \cite{oza}; 0.455~--- 2,
$\triangleleft$ \cite{oza}; 3.27~--- 3, $\vartriangle$ \cite{oza};
9.50~--- 4, $\triangledown$ \cite{oza}, 8.72~--- 5, $+$ \cite{gri};
12.5~--- 6, $\ast$ \cite{gri}; 22.5~--- 7, $\times$ \cite{gri}; 41.7~--- 8,
$\circ$ \cite{gri};  190~--- 9, $\blacktriangle$ \cite{amb};
253~--- 10, $\blacktriangleleft$ \cite{amb}; 307~--- 11,
$\blacktriangledown$ \cite{amb}.} \label{eps33reOGK}
\end{center}
\end{figure}

The temperature dependences of heat capacity of RHS and RDS
crystals along with experimental data \cite{ale}  are depicted in
figure~\ref{cp}.

By a dashed line we show the effective lattice heat capacity
contribution $C_0$, which we estimate as an average difference
$C_\textrm{exp}(T) - \Delta C(T)$. A quantitatively good description of
experiment \cite{ale} is obtained. The calculated value of the
heat capacity jump is  also in a good agreement with experiment.
Deuteration increases the heat capacity in the entire temperature
range.

\begin{figure}[!t]
\begin{center}
\includegraphics[scale=0.98]{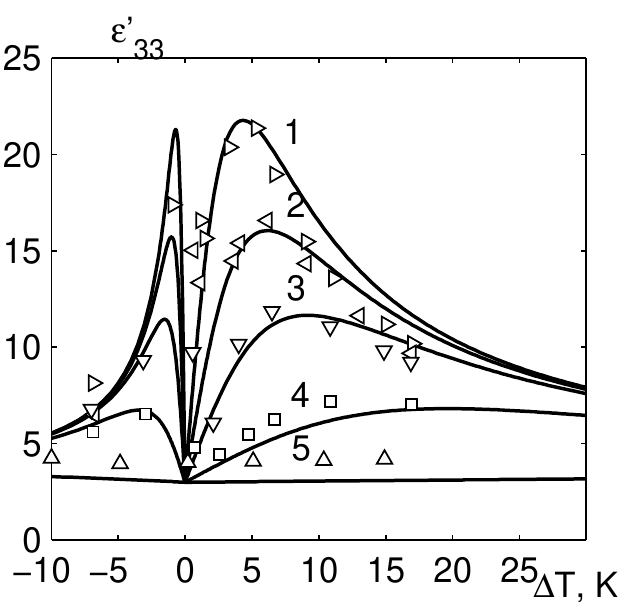}
\hspace{0.5cm}
\includegraphics[scale=0.98]{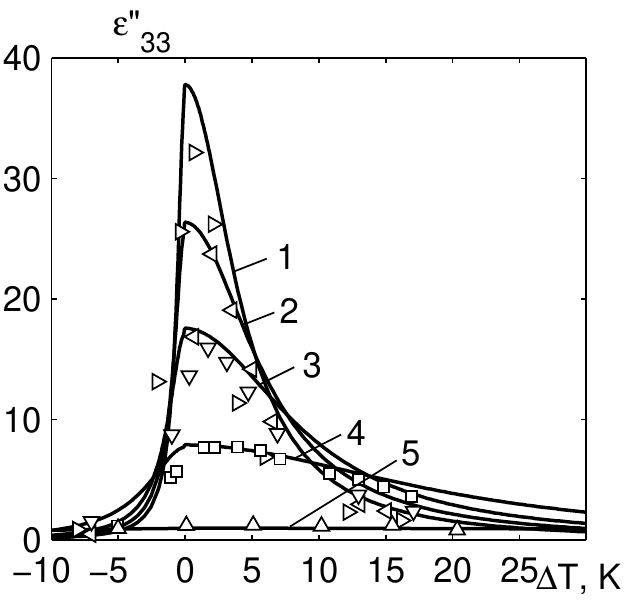}
\end{center}
\vspace{-4mm}
\caption[]{The temperature dependences of the real
$\varepsilon'_{33}$ and imaginary $\varepsilon''_{33}$ parts of
the dynamic dielectric permittivity of Rb(H$_{0.30}$D$_{0.70})$SO$_4$ at different
frequencies $\nu$ (GHz): 8.72~--- 1, $\triangleright$ \cite{gri};
12.5~--- 2, $\triangleleft$ \cite{gri}; 18.72~--- 3,
$\triangledown$ \cite{gri}; 41.70~--- 4, $\square$ \cite{gri}; 330~--- 5, $\vartriangle$ \cite{gri}.} \label{eps33re07G}
\end{figure}

\begin{figure}[!t]
\begin{center}
\includegraphics[scale=0.9]{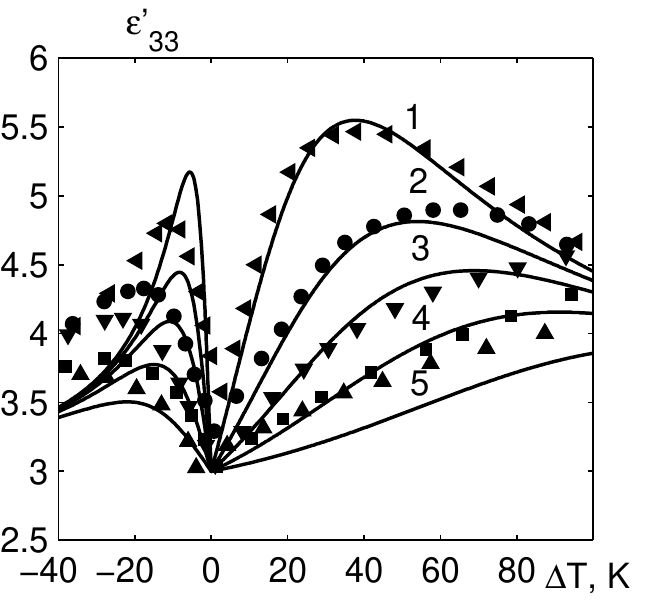}
\hspace{1.5cm}
\includegraphics[scale=0.9]{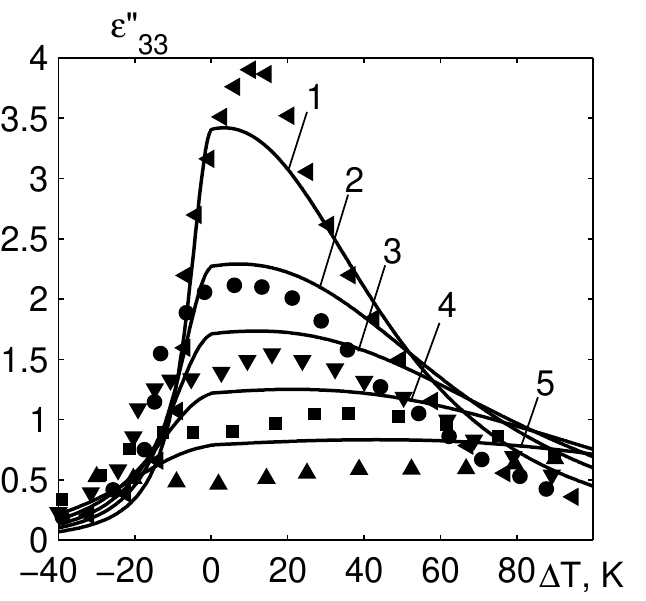}
\end{center}
\vspace{-3mm}
\caption[]{The temperature dependences of the real
$\varepsilon'_{33}$ and imaginary $\varepsilon''_{33}$ parts of
the dynamic dielectric permittivity of RDS at different
frequencies $\nu$ (GHz): 118~--- 1, $\triangleleft$ \cite{amb}; 177~--- 2, $\bullet$ \cite{amb}; 235~--- 3, $\blacktriangledown$ \cite{amb}; 330~--- 4, $\blacksquare$ \cite{amb}; 510~--- 5, $\blacktriangle$ \cite{amb}.}
\label{eps33reKd}
\end{figure}

Figures~\ref{eps33reOGK}--\ref{eps33reKd} show the temperature
dependences of the real $\varepsilon_{33}^{\varepsilon'}(\nu,T)$
and imaginary $\varepsilon_{33}^{\varepsilon''}(\nu,T)$ parts of
the dynamic dielectric permittivity at different frequencies and
compositions of partially deuterated  Rb(H$_{1-x}$D$_x)$SO$_4$
crystals (at $x=0.0$, 0.70, 1.00) along with the experimental data \cite{oza,gri,amb}.

As seen in the figures, the proposed model yields a good
description of experimental data for the
Rb(H$_{1-x}$D$_x)$SO$_4$ crystals over a wide  temperature range
at different frequencies. For all frequencies, at $\Delta T = 0$~K,
the dynamic permittivity $\varepsilon_{33}'(\nu,T)$ has a sharp
minimum, where the permittivity values drop to
$\varepsilon_{33}^{0\varepsilon}$; the minimum width increases
with an increasing frequency. The maximum in the temperature curve of
$\varepsilon_{33}'(\nu,T)$ above $T_\textrm{c}$ lowers down, smears out,
and shifts to higher temperatures  at an increasing frequency. The
dispersion width of the real part of the  permittivity in the
paraelectric phase is wider than in the ferroelectric phase.

\begin{figure}[!t]
\begin{center}
\includegraphics[scale=0.85]{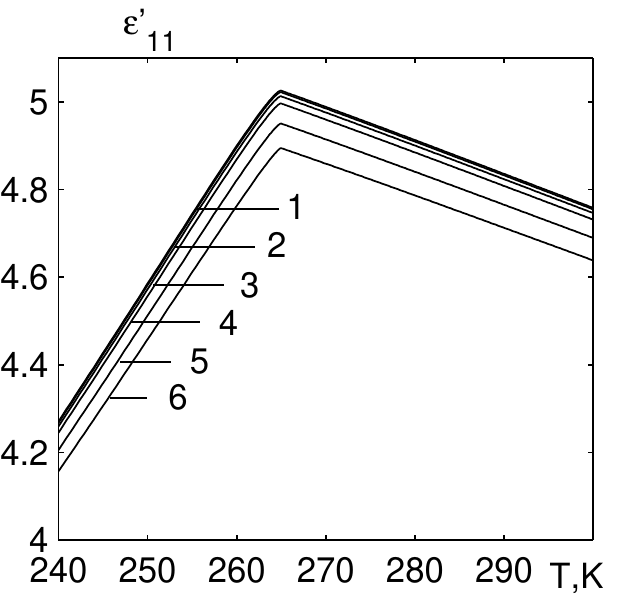}
\hspace{1.5cm}
\includegraphics[scale=0.85]{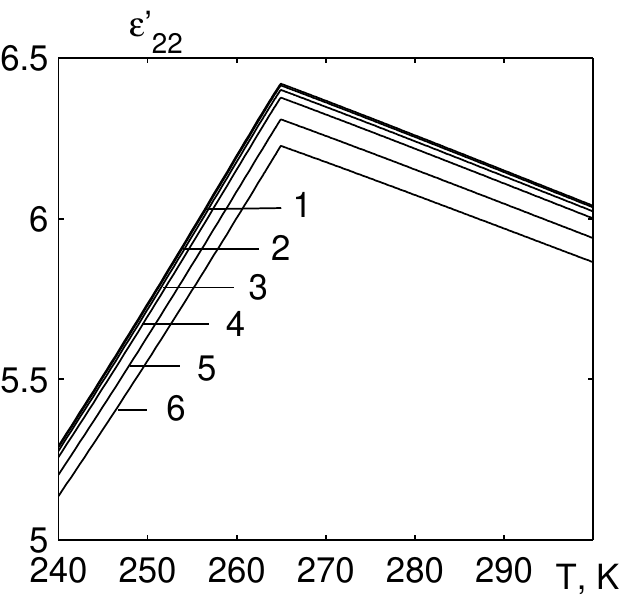}
\caption[]{The temperature dependences of the real
$\varepsilon'_{11}$ and $\varepsilon'_{22}$
parts of the dynamic dielectric permittivity of RDS at different
frequencies $\nu$ (GHz): 0~--- 1; 41.7~--- 2; 78.5~--- 3; 118~--- 4; 190~--- 5;
253~--- 6.} \label{e11re}
\end{center}
\end{figure}

\begin{figure}[!t]
\vspace{-3mm}
\begin{center}
\includegraphics[scale=0.85]{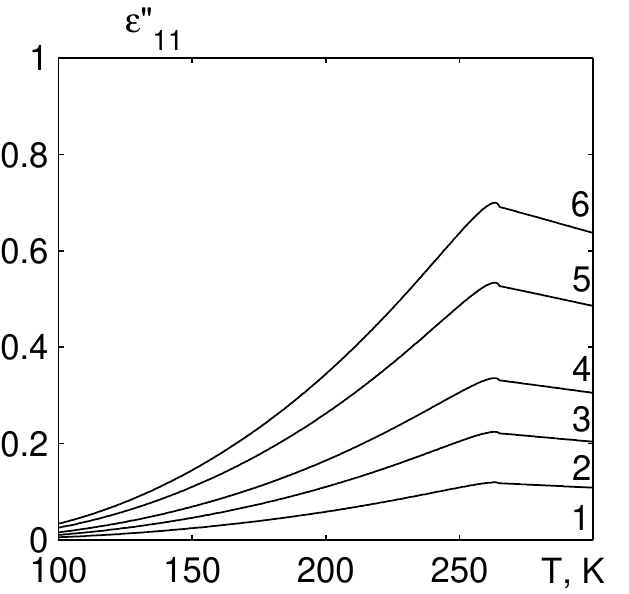}
\hspace{1.5cm}
\includegraphics[scale=0.85]{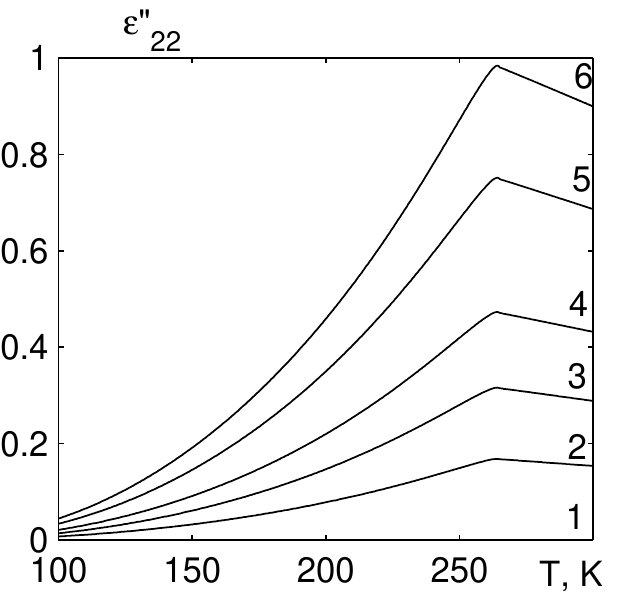}
\end{center}
\vspace{-3mm}
\caption[]{The temperature dependences of the imaginary
$\varepsilon''_{11}$ and  $\varepsilon''_{22}$
 parts of the dynamic dielectric permittivity of RDS at different
frequencies $\nu$ (GHz): 0~--- 1; 41.7~--- 2; 78.5~--- 3; 118~--- 4; 190~--- 5;
253~--- 6.} \label{e22re}
\end{figure}

In figures~\ref{e11re} and \ref{e22re}, we plot the temperature
dependences of the real and imaginary parts of the transverse
dielectric permittivities $\varepsilon'_{11}$,
$\varepsilon''_{11}$ and $\varepsilon'_{22}$, $\varepsilon''_{22}$
of RHS at different frequencies.

\begin{figure}[!t]
\begin{center}
\includegraphics[scale=0.82]{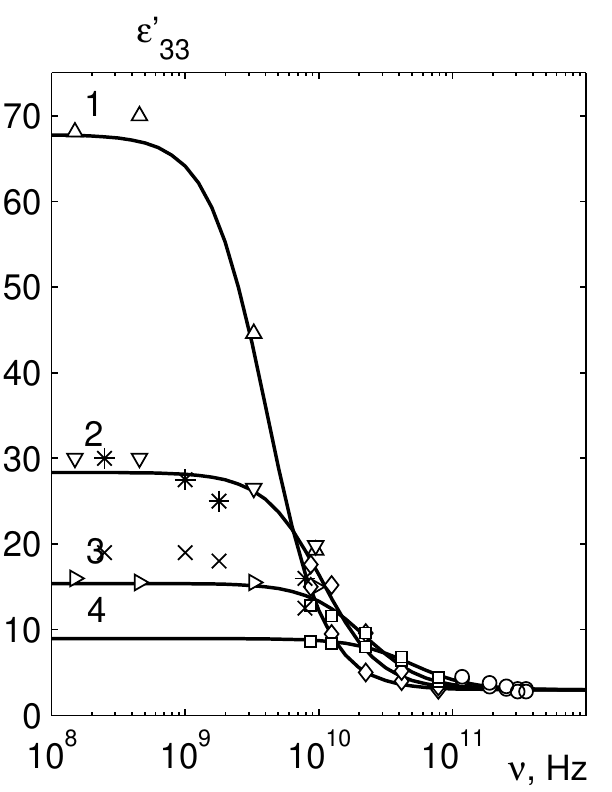}
\hspace{1.5cm}
\includegraphics[scale=0.82]{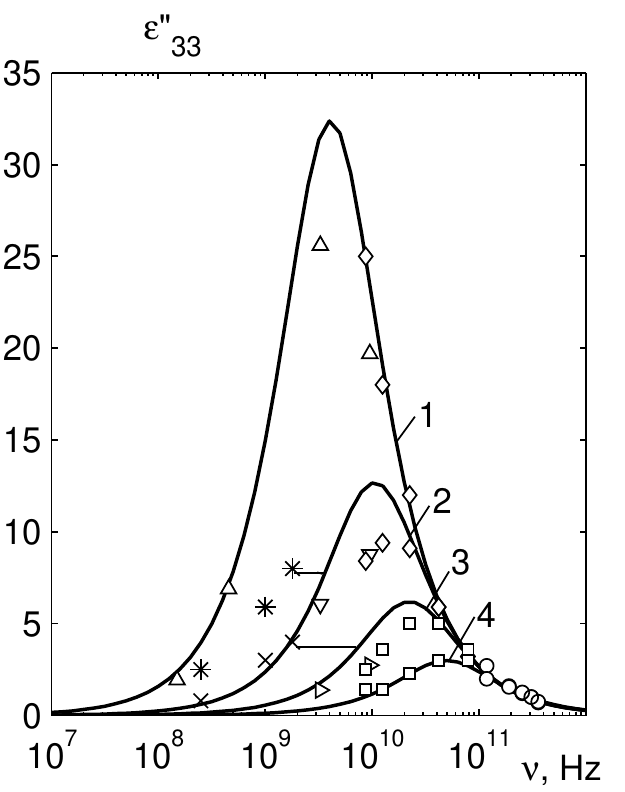}
\end{center}
\vspace{-3mm}
\caption[]{The frequency dependences of $\varepsilon'_{33}(\nu)$
and $\varepsilon''_{33}(\nu)$ for RHS at $\Delta T$: 2~--- 1; 5~---
2; 10~--- 3; 20~--- 4. $\vartriangle$ \cite{oza};
$\bullet$ \cite{kra}; $\circ$ \cite{gri}; $\square$ \cite{amb},
$\times, \ast$ \cite{kra}.} \label{e33re2}
\end{figure}

\begin{figure}[!t]
\begin{center}
\includegraphics[scale=0.85]{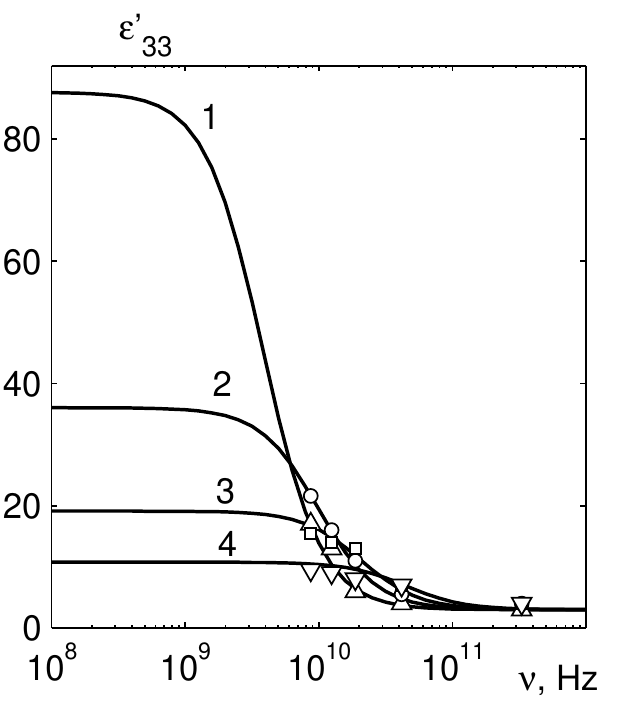}
\hspace{1.5cm}
\includegraphics[scale=0.85]{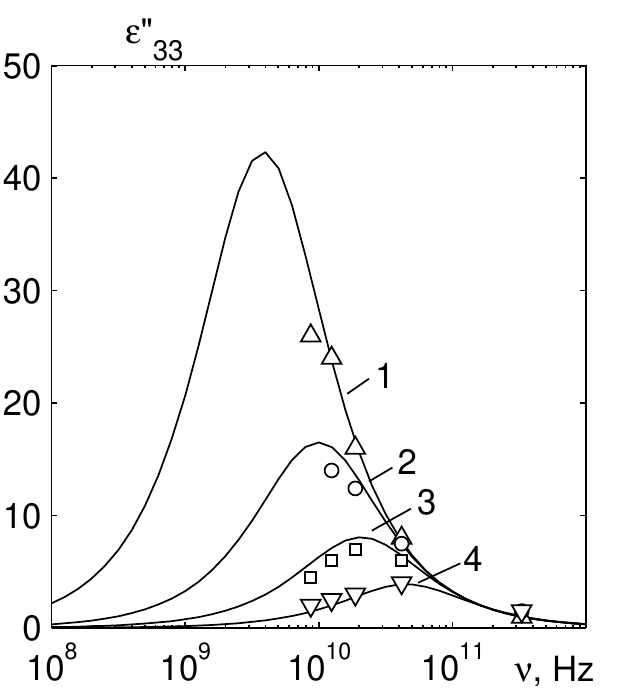}
\end{center}
\caption[]{The frequency dependences of $\varepsilon'_{33}(\nu)$
and $\varepsilon''_{33}(\nu)$ for Rb(H$_{1-x}$D$_x)_2$SO$_4$  at
$x=0.70$ at $\Delta T$: 2~--- 1, $\vartriangle$ \cite{gri}; 5~--- 2,
$\circ$ \cite{gri}; 10~--- 3, $\square$ \cite{gri}; 20~--- 4,
$\triangledown$ \cite{gri}.} \label{e33re07}
\end{figure}

In figure~\ref{e33re2}, we plot the calculated frequency
dependences of $\varepsilon'_{33}(\nu)$ and
 $\varepsilon''_{33}(\nu)$ for RHS at different temperatures $\Delta T= 2$, 5, 10,  20~K and in figure~\ref{e33re07}  for Rb(H$_{1-x}$D$_x)_2$SO$_4$
 with $x=0.70$ along with the experimental points.

As one can see, the theory is in a good agreement with experiment,
except for the  data of \cite{kra} where the permittivity
dispersion was observed at frequencies lower than in \cite{oza,
gri, amb}.  When $\Delta T$ increases, the dispersion of
$\varepsilon_{33}(\nu,T)$  shifts to higher frequencies.

%\clearpage

\section{Conclusions}

In this paper, using the modified four-sublattice  model of a
RbHSO$_{4}$ crystal, with taking into account the piezoelectric
coupling to the $\varepsilon_i$, $\varepsilon_j$ strains, within
the framework of the mean field approach, the theory of the
thermodynamic, dielectric, piezoelectric, elastic, and dynamic
properties of RHS crystals has been developed. A thorough
numerical analysis of the dependences of the calculated
characteristics on the model parameters has been performed.
Optimal sets of these parameters and ``seed'' characteristics for
RHS crystals have been found which enabled us to describe the
available experimental data.

\clearpage

\ukrainianpart

\title{Статистична теорія термодинамічних  та динамічних    властивостей   сегнетоелектрика   RbHSO$_{4}$}
\author{І.Р. Зачек\refaddr{label1}, Р.Р. Левицький\refaddr{label2},
    Я.Й. Щур\refaddr{label2}, О.Б. Біленька\refaddr{label1}}
\addresses{
\addr{label1} Національний університет ``Львівська політехніка'',
вул. С. Бандери, 12, 79013 Львів, Україна
\addr{label2} Інститут фізики конденсованих систем НАН України, вул. І.~Свєнціцького, 1,  79011 Львів, Україна}

\makeukrtitle

\begin{abstract}
Використовуючи модифіковану чотирипідграткову модель RbHSO$_{4}$
шляхом врахування п'єзоелектричного зв'язку з
деформаціями  $\varepsilon_i$, $\varepsilon_4$, $\varepsilon_5$ і $\varepsilon_6$, в
наближенні молекулярного поля  розраховано компоненти вектора
поляризації та тензора статичної діелектричної проникності механічно
затиснутого і вільного кристалів, їх п'єзоелектричні характеристики
і пружні сталі.
 Розраховано також динамічні проникності механічно затиснутого  кристалу RbHSO$_{4}$.
При знайденому наборі параметрів теорії  отримано для цих характеристик задовільний кількісний опис наявних експериментальних даних.
\keywords сегнетоелектрики, діелектрична проникність, п'єзомодулі
\pacs  77.22.Ch, 77.22.Gm, 77.65.-j, 77.80.Bh, 77.84.-s

\end{abstract}

\end{document}